\documentclass[journal=jctcce,manuscript=article,layout=twocolumn]{achemso}
\usepackage{amssymb}
\usepackage{amsmath}
\usepackage{amsbsy}
\usepackage{graphicx}
\usepackage{mathtools}
\usepackage{bm}
\usepackage[symbol]{footmisc}
\usepackage[usenames]{xcolor}

\DeclareMathOperator{\var}{var}
\DeclareMathOperator{\cov}{cov}
\DeclareMathOperator{\diag}{diag}

\def\refpar#1{{(\ref{#1})}}
\DefineFNsymbols*{lamportnostar}[math]{\dagger\ddagger\S\P\|{\dagger\dagger}{\ddagger\ddagger}}
\setfnsymbol{lamportnostar}

\title{Free energy surface reconstruction from umbrella samples using Gaussian process regression}
\author{Thomas Stecher}
\affiliation{Department of Engineering, University of Cambridge, Trumpington Street, Cambridge, CB2 1PZ, UK}
\email{thomas.stecher@tum.de}
\author{Noam Bernstein}
\affiliation{Naval Research Laboratory, Center for Computational Materials Science, Washington, DC 20375, United 
States of America}
\author{G\'abor Cs\'anyi}
\affiliation{Department of Engineering, University of Cambridge, Trumpington Street, Cambridge, CB2 1PZ, UK}
\email{gc121@cam.ac.uk}
\date{\today}

\begin{document}

\begin{abstract}
We demonstrate how the Gaussian process regression approach  can be used to efficiently reconstruct  free energy surfaces from   umbrella sampling simulations. By making a prior assumption of smoothness and taking account of the sampling noise in a consistent fashion, we achieve a significant improvement in accuracy over the state of the art in two or more dimensions or, equivalently, a significant cost reduction to obtain the free energy surface within a prescribed tolerance  in both  regimes of spatially sparse data and short sampling trajectories. Stemming from its Bayesian interpretation the method  provides meaningful error bars without significant additional computation. A software implementation is made available on www.libatoms.org.
\end{abstract}


\section{Introduction}
The free energy is a central quantity in materials science and physical chemistry, governing the macroscopic behavior of a system in the presence of 
thermal fluctuations.  It is defined as a function of a limited number of collective variables, and includes the effects 
of energy as well as entropy, averaged over thermal fluctuations in the remaining degrees of freedom.  The progress 
of any microscopic process occurring in or near equilibrium, including reactions and phase transformations, is 
controlled by the free energy surface in the space of appropriately chosen collective variables.  Many methods exist for computing free energies or their differences, based on 
various aspects and properties of the free energy.

Formally, the free energy is the logarithm of the marginal probability density of the system in equilibrium corresponding to the 
chosen thermodynamic ensemble. A crude and na\"ive  approach would therefore be to  estimate the probability density (e.g.\ 
by constructing histograms) from samples obtained in a direct molecular dynamics (MD) or Monte Carlo (MC) 
simulation. This  is almost never practical in scientifically interesting cases due to the presence of large barriers and metastable states which cannot 
be representatively sampled directly on the time scale of a typical direct simulation. One solution to this problem is to 
restrain the sampling to specific areas of interest by altering the Hamiltonian through an additional potential term, 
often called a {\em bias}. This is the umbrella or non-Boltzmann sampling approach and goes back to Torrie and 
Valleau\cite{umbrella}, who showed how to recover the unbiased probability distribution from the biased one.

In this paper we focus on the question of how to efficiently use the data coming from a set of such biased simulations. Traditional methods include the Weighted Histogram Analysis Method (WHAM)
\cite{WHAM1,WHAM2} and the Umbrella Integration (UI)\cite{UI1,UI2} technique, including its multidimensional 
variant\cite{Maragliano:2008hw, Kastner:2009ey} which uses least squares fitting of radial basis functions (LSRBF)
\cite{Buhmann:2000tc} to reconstruct the free energy surface. The former takes direct advantage of the  relation 
between the free energy and the marginal probability distribution of a system; the latter - like thermodynamic 
integration\cite{kirkwood} on which it is based - reconstructs the free energy from the mean (i.e.\ thermodynamically 
averaged) force in each biased simulation, which under suitable conditions is the negative gradient of the free energy.
Recently a number of maximum-likelihood approaches operating on the probability distribution, such as multistate Bennett acceptance ratio (MBAR)\cite{Shirts_JCP_2008}, which has been shown to be equivalent to a binless WHAM\cite{Tan_JCP_2012}, and variational free energy profile (vFEP)\cite{Lee_JCTC_2013,Lee_JCTC_2014}, have
been proposed that also operate on the histogram and go some way to eliminate the shortcomings of WHAM.  

Working with more than one collective coordinate significantly adds
to the challenge of free energy surface reconstruction. The most obvious
 problem, which gets exponentially harder as the
number of dimensions increases, is exploration, i.e.\ the problem
of ensuring that the entire region of interest is sufficiently
sampled. The principal difficulty is a chicken-and-egg problem:\ before the free energy surface is reconstructed, it is not possible to know which regions of collective variable space correspond to low free energies and therefore are relevant. 

Some approaches, such as metadynamics~\cite{Laio:2002wm}
and the adaptive biasing force (ABF) method~\cite{Darve:2001ce,Darve:2008iw},
combine exploration and reconstruction by gradually building up an
approximate bias potential that is designed to converge to the
negative of the free energy and thus counteract the
natural occupancy bias of the Boltzmann distribution towards low
free energy regions.  Metadynamics does this based on where the
system has been, i.e. its observed probability density, while ABF uses
sampled free energy gradients.
Another approach is to separate exploration from reconstruction,
as in the ``single-sweep''
method.\cite{Maragliano:2008hw} In this approach, temperature-accelerated
molecular dynamics\cite{Maragliano:2006ei} -- an extension
of adiabatic free energy dynamics\cite{Rosso:2002gs,
Abrams:2008jl}  -- is used in a first step to identify a small number of
points in the ``important regions of the free energy
landscape''.\cite{Maragliano:2008hw} At these points free energy
gradients are  calculated using independent standard umbrella
sampling simulations and then, in the second step, the free energy surface is
reconstructed from these gradients using LSRBF. Recently, Tuckerman
and co-workers\cite{Chen:2012js} proposed what amounts to a synthesis
of metadynamics and the single-sweep method, combining dynamics at
a higher temperature in the (adiabatically decoupled) collective
variables with an on-the-fly construction of a bias potential to
steer the system away from regions in collective-variable space
already visited. The free energy is again reconstructed from
measurements of its gradients in umbrella simulations and least
squares fitting (but using fewer basis functions compared to the original "single sweep" method).

Gaussian mixtures umbrella sampling (GAMUS) is another recently published combined exploration-reconstruction method, in which the  probability distribution is modelled by a sum of Gaussians whose parameters are determined using the Expectation-Maximisation algorithm.\cite{Maragakis_JPCB_2009} The authors warn that it is primarily useful for exploration, and not for determining the free energy profile accurately.

In the present work, we do not tackle the first task of exploration; rather, we seek
to showcase the Gaussian process regression (GPR) framework for performing the second task of free energy
reconstruction from independent umbrella simulations.  In the one dimensional case only, exploration
rarely preoccupies the modeller: usually a simple 
uniform grid over the range of interest suffices. 
In more than one dimension we wish to demonstrate  how GPR
compares with LSRBF and vFEP. LSRBF is known to have problems with
large condition numbers in cases where the data density is
high,\cite{Buhmann:2000tc,Maragliano:2008hw} and both it and vFEP will be shown below to be rather intolerant of noise in the data.  Because the simulations needed for exploration and collection
of free energy gradients are so computationally expensive, the {\em
post hoc} reconstruction of the free energy surface is a negligible
part of the overall computational cost for all the methods we investigate and therefore we omit discussion of computational cost directly associated with the free energy surface reconstruction.

Bayesian statistics is a natural framework for function fitting, allowing us not only to express our prior beliefs about 
the smoothness of the free energy function in terms of a prior probability distribution defined on a Hilbert space of a 
large class of possible functions, but also to treat consistently the often significant sampling noise arising from the 
limited time span of the  trajectories. More specifically, GPR is an ideal 
technique for our purposes, because the evaluation of the reconstructed free energy function is analytically tractable, and one does 
not have to perform further sampling  just to evaluate the reconstructed function, as is often the case with more 
complex Bayesian techniques. The GPR framework allows one to incorporate all the useful data from the MD 
trajectories in a consistent fashion, including both histograms and measurements of free energy derivatives.  Our exposition of the theory behind the GPR is 
nevertheless rather basic, because there are excellent text books on the subject. Our purpose is to illustrate the kinds of benefits these types of methods can offer for free 
energy reconstruction, and more sophisticated reconstruction algorithms are possible based on the available 
statistics and machine learning  literature. 

Indeed the use of Gaussian processes for prediction and interpolation problems is widespread. While the basic theory came from time series analysis\cite{Kolmogorov:1941uk,Wiener:1949uo}, it became 
popular in the field of geostatistics under the name ``kriging''\cite{Matheron:1973tn}. O'Hagan\cite{OHagan:1978tx} 
generalised the theory, while Williams and Rasmussen\cite{Williams:1996ve, Rasmussen:2005ws} brought the 
approach to machine learning, thus disseminating it to a wider community. In the field of physical chemistry, Bart\'ok 
\textit{et al.} recently used it to create highly accurate potential energy surfaces for selected materials by 
interpolating \textit{ab initio} energies and forces\cite{Bartok:2010wd,Bartok:2013} , while Rupp \textit{et al.} 
interpolated the atomisation energies of a wide range of molecules,\cite{Rupp:2012} and Handley \textit{et al.} interpolated electrostatic multipoles.\cite{Handley:2009}

The paper is structured as follows: In Sec.~\ref{sec:background} we briefly review umbrella sampling and the 
WHAM and UI techniques for free energy reconstruction to define our notation, and then  describe the formalism of  Gaussian 
process regression. In Sec.~\ref{sec:BRUSH} we propose a free energy reconstruction method that combines 
umbrella sampling, histogram analysis and GPR. In Sec.~\ref{sec:GPUI} we introduce an alternative method which 
uses GPR  to reconstruct a free energy profile from the mean forces obtained from the same data, before showing in 
Sec.~\ref{sec:hybrid} how in fact the two strategies can be combined in a consistent manner. We discuss error estimates in Sec.~
\ref{sec:errorbars} but defer the detailed derivations to the Appendix. 
We compare the numerical performance of our approaches to that of WHAM,  UI, LSRBF and vFEP in Sec.~\ref{sec:performance}.


\section{Background Theory}\label{sec:background}


\subsection{Umbrella Sampling}\label{sec:UmbSamp}
Umbrella sampling\cite{umbrella} provides the means to obtain and use samples from biased MD or MC simulations 
in free energy reconstructions and related problems. Biased sampling is necessary whenever the system of interest 
has appreciable energetic or entropic barriers. In such cases the transitions between metastable states separated 
by these barriers occur over a much longer time scale than that of the fast molecular vibrations that put a strong 
lower bound on the time resolution of the simulation, and a corresponding constraint on the total simulation time 
possible given fixed computational resources. It would therefore not be possible to obtain an accurate 
representation of the global probability distribution in configuration space without biasing. 

In the following we write $\beta=1/k_{\mathrm{B}}T$ for the inverse temperature, $\mathbf{q}$ for the configurational 
degrees of freedom of the system, $x=s(\mathbf{q})$ for the collective variable of interest, $V(\mathbf{q})$ for the 
potential energy and 
\begin{equation}
Z=\int e^{-\beta V(\mathbf{q})}d\mathbf{q}
\end{equation}
for the configurational partition function of the system.
Biasing is achieved by the addition of a restraining potential $w(s(\mathbf{q}))$, often chosen to be harmonic. This 
changes the original, unbiased probability density of the system
\begin{equation}
P(x)=\frac{1}{Z}\int e^{-\beta V(\mathbf{q})}\delta(s(\mathbf{q})-x)d\mathbf{q}
\end{equation}
to the biased distribution
\begin{align}
P^b(x)&=\frac{1}{Z^b}\int e^{-\beta [V(\mathbf{q})+w(s(\mathbf{q}))]}\delta(s(\mathbf{q})-x)d\mathbf{q}\nonumber\\
&=\frac{1}{Z^b}e^{-\beta w(x)}\int e^{-\beta V(\mathbf{q})}\delta(s(\mathbf{q})-x)d\mathbf{q},
\end{align}
where  
\begin{equation}\label{eq:biaspart}
Z^b=\int e^{-\beta [V(\mathbf{q})+w(s(\mathbf{q}))]}d\mathbf{q}
\end{equation}
is the configurational partition function of the biased system.
The unbiased distribution can thus be reconstructed from the biased distribution as\cite{umbrella}
\begin{equation}
P(x)=\frac{Z^b}{Z}e^{\beta w(x)}P^b(x).
\end{equation}
However, the partition function ratio, given by the following expressions, cannot be evaluated directly because
contributions from poorly sampled 
areas dominate the average.
\begin{align}
\frac{Z^b}{Z}=\frac{\int e^{-\beta[V(\mathbf{q})+w(s(\mathbf{q}))]}d\mathbf{q}}{\int e^{-\beta V(\mathbf{q})}d\mathbf{q}}
&=\langle e^{-\beta w(s(\mathbf{q}))}\rangle\label{eq:partition1}\\
=\frac{\int e^{-\beta[V(\mathbf{q})+w(s(\mathbf{q}))]}d\mathbf{q}}{\int e^{-\beta [V(\mathbf{q})+w(s(\mathbf{q}))]}
e^{+\beta w(s(\mathbf{q}))}d\mathbf{q}}
&=\frac{1}{\langle e^{\beta w(s(\mathbf{q}))}\rangle_b},\label{eq:partition2}
\end{align}
where $\langle\cdots\rangle$ denotes a canonical average of the unbiased and $\langle\cdots\rangle_b$ a 
canonical average of the biased system. Eq.~\eqref{eq:partition1} will suffer from the usual problems of unbiased sampling, i.e.\ 
convergence will only be achieved on an appreciable time-scale if the minima of $V$ and $w$ are close to each 
other, a prerequisite which defies the point of biasing. Eq.~\eqref{eq:partition2} on the other hand will record the most significant 
contributions to the average where the bias potential is large and therefore the density of the biased distribution is
low.


\subsection{WHAM}
When using methods that employ only one global (perhaps complicated) bias potential, such as metadynamics
\cite{Laio:2002wm} or the adaptive biasing force (ABF) method\cite{Darve:2001ce}, the unknown normalisation 
constant $Z^b/Z$ is not needed: the unbiased probability distribution is simply reconstructed up to a multiplicative 
constant and, if required, renormalised \textit{a posteriori} so that it integrates to unity. However, if samples come 
from several simulations, each with its own bias, as is the usual case in umbrella sampling with many umbrella 
windows, the unknown normalisation constant will differ from window to window. WHAM\cite{WHAM1,WHAM2} 
solves this problem using an iterative procedure, effectively matching the free energy, $F(x)=-\frac{1}{\beta}\ln P(x)$, 
in the overlap regions between each pair of windows. This necessitates a simulation setup which ensures good 
sampling of these overlap regions.

More specifically, WHAM constructs a global histogram with a set of bins spanning all umbrella simulations. Using 
$N_{\theta}$ for the total number of samples collected in window $\theta$, the expected number of bin counts, $
\langle n_{\theta i}\rangle$, in the \(i\)th bin in the biased simulation corresponding to window \(\theta\) can be written in terms of the unknown unbiased bin 
probabilities $P_i$ as
\begin{equation}
\langle n_{\theta i}\rangle = N_\theta \frac{P_i c_{\theta i}}{\sum_i P_i c_{\theta i}},\label{eq:binexpect}
\end{equation}
where $c_{\theta i}$ represents the multiplicative bias value in window $\theta$ against bin $i$, approximated by the value 
of the bias at the centre, $x_i$, of each bin, i.e.
\begin{equation}
c_{\theta i}=e^{-\beta w_{\theta}(x_i)}.
\label{eq:cthetai_approx}
\end{equation} 
The sum in the denominator in equation \eqref{eq:binexpect} is an approximation to the biased partition function 
associated with the biased simulation, corresponding to window $\theta$,
\begin{equation*}
Z^b_{\theta}=\sum_i c_{\theta i}P_i.
\end{equation*}
In order to determine the unknown unbiased probabilities $P_i$, equation \eqref{eq:binexpect} is first summed over 
the windows, 
\begin{equation*}
\sum_\theta \langle n_{\theta i}\rangle = \sum_\theta N_\theta P_i c_{\theta i} [Z^b_\theta]^{-1},
\end{equation*}
and then rearranged to express $P_i$,
\begin{equation}
P_i=\frac{\sum_\theta \langle n_{\theta i}\rangle}{\sum_{\theta}N_{\theta}[Z^b_\theta]^{-1}c_{\theta i}}.
\end{equation}

In order to use it to recover unbiased bin probabilities from observed bin counts, we replace the expectation in the numerator by the sum of actual observed bin counts, $M_i = \sum_\theta 
n_{\theta i}$,  and then solve for the set of $P_i$-s iteratively until self-consistency is achieved. It has been shown
\cite{Bartels:1997wl,Bartels:2000fk} that the solution thus obtained is equivalent to maximising the likelihood of 
jointly observing the bin counts $\{M_i\}$ given underlying bin probabilities $\{P_i\}$. Indeed, Zhu and Hummer
\cite{Zhu:2011fo} have shown that maximising the likelihood directly leads to a more efficient algorithm compared to the 
traditional iterative procedure. 

Equation~\eqref{eq:binexpect} can be used to directly illuminate why the distributions in the neighbouring umbrella 
windows must overlap significantly in order for WHAM to work. With the anticipation of taking logarithms in the next 
step, we define $y_{\theta i}$ and $\tilde{y}_{\theta i}$ by writing the bin count values as
\begin{align}
\frac{n_{\theta i}}{N_\theta} &= e^{-\beta y_{\theta i}},\label{eq:ydef}\\
\frac{\langle n_{\theta i} \rangle}{N_\theta} &= e^{-\beta \tilde{y}_{\theta i}},
\end{align}
so that $y_{\theta i}$ corresponds to the actually observed values and $\tilde{y}_{\theta i}$ corresponds to 
expectations. If we denote the unknown unbiased free energy at bin $i$ by $F_i =  - (\ln P_i)/\beta$ and the 
logarithm of the unknown bin partition function by $f^b_\theta = \ln Z^b_\theta$, then the logarithm of equation~
\eqref{eq:binexpect} becomes a set of equations, one for each $i$ and $\theta$, 
\begin{equation}
F_i  = \tilde{y}_{\theta i} - w_\theta(x_i) + f^b_\theta \approx y_{\theta i} - w_\theta(x_i) + f^b_\theta,
\label{eq:free_e_overlap}
\end{equation}
where the value of each $F_i$ has to be consistent for all $\theta$, and the approximation results from replacing the 
expectation $\tilde{y}_{\theta i}$ by the observed $y_{\theta i}$.  In addition to this approximation,  if $n_{\theta i}$ 
happens to be zero for some particular umbrella $\theta$ and bin $i$, its logarithm $y_{\theta i}$ would be 
undefined, and the instance of Eq.~\eqref{eq:free_e_overlap} corresponding to that $\theta$ and $i$ combination is 
therefore ignored.  The coupling between different umbrellas occurs when a given bin has non-zero contributions 
from at least two umbrellas, and so all instances of Eq.~\eqref{eq:free_e_overlap} with the same $i$ and different $
\theta$ must hold with the same value of $F_i$.  The task of finding a consistent set of values for all $F_i$ and 
$f^b_\theta$ is equivalent to the self-consistency iteration in the conventional formulation of WHAM.  However, if 
there exists an umbrella that has non-zero bin counts $n_{\theta i}$ only for bins that do not have any counts from 
any other umbrellas, there is no need for consistency, and that set of bins is decoupled from the rest.  The free 
energy of that set of bins is therefore ill defined: an arbitrary constant can be added to their $F_i$ and $f^b_\theta$ 
without disturbing any $F_i$ or $f^b_\theta$ for any other bins or umbrellas, thus defeating the whole purpose of 
generating a globally valid reconstruction.  


\subsection{Umbrella Integration}
An alternative approach, umbrella integration, is based on measuring the gradient of the free energy obtained from a 
set of simulations using strictly harmonic bias potentials
\begin{equation}
w_{\theta}(x)=\frac{1}{2}\kappa_{\theta}(x-x_{\theta})^2,
\end{equation}
where $\kappa_{\theta}$ is the strength of the bias potential and $x_{\theta}$ the window centre.
In each window, there is a collective variable value where the restraining force from the bias potential is exactly 
balanced by the free energy derivative, thus creating a stationary point in the biased distribution, namely its mode, $
\hat{x}_{\theta}$.\cite{UI1} Therefore, at the collective variable value corresponding to the mode,  the free
energy gradient is equal to minus the restraint gradient.
In practice, however, the mode of the biased distribution is approximated by its mean,  $\bar{x}_{\theta}$, a quantity 
which can be estimated much more accurately from a limited amount of data than the mode, giving the following 
estimated free energy gradients\cite{UI1}
\begin{align}
\left.\frac{\partial F(x)}{\partial x}\right|_{\hat{x}_{\theta}}&=
- \left.\frac{\partial w_\theta}{\partial x}\right|_{\hat{x}_{\theta}}\nonumber\\
&\Downarrow\quad \hat{x}_{\theta}\approx \bar{x}_{\theta}\nonumber\\
\left.\frac{\partial F(x)}{\partial x}\right|_{\bar{x}_{\theta}}&\approx
- \left.\frac{\partial w_\theta}{\partial x}\right|_{\bar{x}_{\theta}}=-\kappa_{\theta}(\bar{x}_{\theta}-x_{\theta})
\label{eq:UIder}
\end{align}

The  mode exactly equals the mean  if the biased distribution is unimodal and symmetric,\cite{UI1} and the 
approximation will be good if the harmonic restraining potential is sufficiently stiff. However, the variance of the estimator in~\eqref{eq:UIder} scales
like $O(\kappa)$, so stronger biases lead to more statistical noise. Also,
 using very stiff 
bias potentials necessitate the use of short time-steps in the simulation
and might introduce high barriers in the 
space perpendicular to the collective variable, thus making the sampling
very inefficient. 
   
Working with free energy derivatives rather than absolute free energies allows UI to avoid the problem of the 
unknown  offsets or equivalent normalization constants that WHAM encounters. In one dimension it is simple to integrate
the estimated gradients numerically to obtain the free energy profile. In more than one dimension one needs to ensure that the the free energy surface is independent of the path of integration. One solution to this problem is provided by least squares fitting, which we discuss in Sec.~\ref{sec:2+Dtheory}. 

\subsection{Gaussian Process Regression}\label{sec:GPR}
In this section we outline the most important aspects of GPR, the main computational technique underlying our 
proposed method, and define notation.  An in-depth introduction to GPR can be found in the textbooks\cite{Rasmussen:2005ws,MacKay:2003wc}. In the most basic case, we have a number of noisy scalar readings, collected into a vector $\mathbf{y} = \{y_i\}$, of an 
unknown function $f$, at a certain number of locations $\mathbf{x} = \{x_i\}$, and we wish to predict the function value, 
$f(x^*)$, at a new location $x^*$. GPR is a Bayesian method: it combines a prior probability distribution on the 
function values with the data through the likelihood of  measuring the observed data given the function. Formally, 
\begin{align*}
p(f(x^*)|\mathbf{y}) &= p(f(x^*) ~\mathrm{and}~ \mathbf{y})/p(\mathbf{y})\\
&\propto \int d\mathbf{f}(\mathbf{x}) p(f(x^*) ~\mathrm{and}~ \mathbf{f}(\mathbf{x})) p(\mathbf{y}|\mathbf{f}(\mathbf{x})),
\end{align*}
where the elements of the vector $\mathbf{f}(\mathbf{x}$) are the (unknown) values of the underlying function $f$ at 
the measurement locations $\mathbf{x}$. Note that the normalising constant of the posterior probability distribution 
$p(\mathbf{y})$ is typically not calculated explicitly, and has been omitted on the second line. The first term of the product in the integral  corresponds to the prior distribution 
on $f$, and thus does not depend on the values of the observed data, only on its location, while the second term is 
the likelihood which encodes information about the noise inherent in our measurement. The output of the GPR is the 
posterior distribution on the left hand side, corresponding to the prediction at an arbitrary new location $x^*$. We will 
interpret the maximum of the posterior distribution as corresponding to the most likely reconstructed function value 
given the data and the prior.

In GPR the prior distribution is a Gaussian process, which describes a distribution over functions and can loosely be 
thought of as an infinite-dimensional multivariate Gaussian distribution or, more formally, a collection of random 
variables, any finite number of which have a joint Gaussian distribution\cite{Rasmussen:2005ws}. Just as a 
multivariate Gaussian distribution is defined by a mean vector and covariance matrix, a Gaussian process is defined 
by a mean function $m(x)$ and a covariance function (or kernel) $k(x,x^\prime)$. Throughout this paper we will take 
the prior mean function to be zero, so the prior is,
\begin{equation}
p(\mathbf{f}(\mathbf{x})) = (2\pi)^{-n/2}|K(\mathbf{x})|^{-1/2} \exp[- \frac12 \mathbf{f}^T K(\mathbf{x}) \mathbf{f}],
\label{eq:prior}
\end{equation}
where  $n$ is the number of data points, $K(\mathbf{x})$ is the covariance matrix, and $\left|K(\mathbf{x})\right|$ is
its determinant.  The entries $K_{ij}=k(x_i,x_j)$ give the covariance between
the values of the function evaluated at $x_i$ and at $x_j$. We discuss the specification of the prior 
covariance function in more detail in the next section but, loosely, the value of the covariance between two locations indicates how close the corresponding function values are expected to be.

The measurement noise is also assumed to be joint Gaussian, i.e.\ the likelihood takes the form,
\begin{multline}\label{eq:GPRlikelihood}
p(\mathbf{y}|\mathbf{f}(\mathbf{x}))=(2\pi)^{-n/2}|\Sigma_y|^{-1/2}\times\\
\times \exp\left[-\frac{1}{2}(\mathbf{y}-\mathbf{f}(\mathbf{x}))^T\Sigma_y^{-1}(\mathbf{y}-\mathbf{f}(\mathbf{x}))\right],
\end{multline}
where $\Sigma_y$ is the noise covariance matrix and $|\Sigma_y|$ is its 
determinant. (Note that if the measurements are independent, this matrix is diagonal with the noise variance of the measurements in the diagonal elements of $\Sigma_y$.) Under these assumptions the posterior distribution turns out to be also a 
Gaussian process with mean and covariance functions given by\cite{Rasmussen:2005ws}
\begin{align}
\bar{f}(x^*)&=\mathbf{k}^T(x^*,\mathbf{x})(K(\mathbf{x})+\Sigma_y)^{-1}\mathbf{y},\label{eq:GPpostmean}\\
\cov(f(x_1^*),f(x_2^*))&=k(x_1^*,x_2^*)-\mathbf{k}^T(x_1^*,\mathbf{x})\cdot\nonumber\\ &\quad \cdot(K(\mathbf{x})+\Sigma_y)^{-1}\mathbf{k}
(x_2^*,\mathbf{x}), \label{eq:GPpostvar}
\end{align}
where $\mathbf{k}(x^*,\mathbf{x})$ is the vector composed of 
covariance function values $\{k(x^*,x_i)\}$. For notational convenience and to keep our equations compact and readable 
later on, we shall often suppress the dependence on the observation locations $\mathbf{x}$ and write $K_y$ for $K+
\Sigma_y$, in line with Ref.~\citenum{Rasmussen:2005ws}.  The posterior mean, $\bar{f}(x^*)$, (which coincides 
with the posterior mode and median) is often reported as the function reconstruction of the GPR model, while the 
diagonal elements of the posterior covariance function provide error bars.

According to Eq.~\eqref{eq:GPpostmean} the prediction $\bar{f}(x^*)$ is simply a linear combination of kernel 
functions centred on the data points:\cite{Rasmussen:2005ws}
\begin{align}
\bar{f}(x^*)&=\sum_{i=1}^n b_i k(x^*,x_i),\label{eq:GPRansatz}\\
\mathbf{b}&=(K(\mathbf{x})+\Sigma_y)^{-1}\mathbf{y}\nonumber
\end{align}
While this is not unlike a least-squares 
fit\cite{Buhmann:2000tc}, it differs in the way the coefficient vector $\mathbf{b}$ is calculated. In 
particular, through the presence of $\Sigma_y$ GPR takes account of the noise associated with the data and, through the prior, of the similarity of data points, while the radial basis approach in its 
most basic form attempts to fit the data -- regardless of its noise -- exactly.
As can be seen from Eq.~(\ref{eq:GPRansatz}), the covariance functions of
GPR are analogous to the basis functions used in LSRBF (see Sec.~\ref{sec:2+Dtheory}), but the covariance  has a statistical meaning, and this will play a significant role in what follows, in particular in how any free parameters are
set.  

The numerical implementation of GPR in the simplest case of noisy readings of function values is straightforward.  
From the sample positions, values, and noise, one pre-computes a single 1-dimensional array $\mathbf{b}$:\medskip

\framebox{\parbox{8cm}{
\begin{enumerate}
   \item Create an array of data positions $\{x_i\}$ and an array of corresponding data values $\{y_i\}$.
   \item Compute the matrix $K$ from data positions with $K_{ij} = k(x_i, x_j)$. 
   \item Evaluate the noise variance matrix $\Sigma_y$ for the data values. 
   \item Compute the array\\$b_i = \sum_j {\left(K + \Sigma_y\right)^{-1}_{ij} y_j}$.
\end{enumerate}
}}

\medskip

\noindent Evaluating the prediction of the model $\bar{f}(x^*)$ at an arbitrary position $x^*$ just requires that one\medskip

\framebox{\parbox{8cm}{
\begin{enumerate}
   \item Compute an array of covariances\\ $k^*_i = k(x^*,x_i)$.
   \item Compute the prediction\\ $\bar{f}(x^*) = \sum_i {k^*_i b_i}$
\end{enumerate}
}}


\subsection{Covariance Functions and Hyperparameters: Prior Choices}\label{sec:prior}
 The choice of covariance 
function determines how smooth the Gaussian process reconstruction will be. Many covariance functions have been proposed for the purpose of Gaussian process regression.\cite{Rasmussen:2005ws}
We have found the common choice of the squared exponential (SE) covariance function,
\begin{equation}\label{eq:SE}
k(x_1,x_2)=\sigma_f^2\exp\left(-\frac{(x_1-x_2)^2}{2 l^2}\right),
\end{equation}
to suffice for our purposes in this paper. It is infinitely differentiable an thus ensures a particularly smooth function 
reconstruction. We refer to it as a squared exponential (rather than Gaussian) in line with Ref.~
\citenum{Rasmussen:2005ws} in order to emphasise that this choice is not a prerequisite for Gaussian process 
regression and that other options are available. The name ``Gaussian'' in GPR refers to the probability  distributions, rather than this choice of covariance function.   

If we have the prior knowledge that the underlying function is periodic (e.g.\ the free energy corresponding to a 
dihedral angle), this should be reflected in the covariance function used. MacKay\cite{MacKay:1998wg} has shown 
how this can be achieved for the SE and other covariance functions. The (periodic) variable $x$ is first mapped onto 
a circle as $\mathbf{u}(x)=(\cos(x),\sin(x))$ and the covariance function then applied to $\mathbf{u}(x)$. For the SE 
covariance function one obtains
\begin{align}\label{eq:periodiccov}
k_{2\pi}(x_1,x_2)&=\sigma_f^2\exp\left(-\frac{|\mathbf{u}(x_1)-\mathbf{u}(x_2)|^2}{2l^2}\right)\nonumber\\
&=\sigma_f^2\exp\left(-\frac{2\sin^2\left(\frac{x_1-x_2}{2}\right)}{l^2}\right).
\end{align}
We use this covariance function to treat dihedral angles in this paper.  In two or more dimensions, as in 
Secs.~\ref{sec:two_dimensions} and \ref{sec:four_dimensions}, these expression are generalized to
\begin{align}
k_{2\pi}(x_1,x_2)=\,& \sigma_f^2\exp\left(-2 \sum_\alpha \frac{\sin^2\left(\frac{x_{1,\alpha}-x_{2,\alpha}}{2}\right)}{l_\alpha^2}-\right.\nonumber\\
& \left. \sum_\beta\frac{\left(x_{1,\beta}-x_{2,\beta}\right)^2}{2 l_\beta^2}\right) ,
\end{align}
where $\alpha$ and $\beta$ are indices over periodic and non-periodic collective coordinates, respectively, each of which can have its own length scale.

Finally, we  note that these covariance functions contain a number of free parameters (commonly referred to as 
``hyper-parameters'' in the machine learning literature), such as the length scale $l$ of the function to be inferred and 
the prior function variance $\sigma_f^2$, i.e.\ the expected variance of the function as a whole. Our view is that 
these are strictly part of the prior and should be informed by our prior knowledge of the system we investigate, rather than 
measured from the data itself. E.g. the meaning of \(l\) is the
distance scale over which values of the unknown function are expected to
become uncorrelated.
 The smoother we expect the reconstructed function to be, the larger  \(l\)
we should choose. Arguably in most free energy calculations
we have a very good idea about the length 
and energy scales over which the free energy typically changes (e.g.\ the
length and strength of a chemical bond, or simply $\pi$ in the case of a
bond or dihedral angle). Furthermore, we note that the effect of these  choices on the reconstruction will become weak with increasing 
amounts of data. The same is not true for the noise covariance,  $\Sigma_y$, since more precise measurements  
correspond to lower values and this relationship needs to be maintained.


\section{Bayesian Free Energy Reconstruction from Histograms}\label{sec:BRUSH}
We first illustrate the theory by the reconstruction of free energy profiles from histogram data. The basic idea is the 
following: For every umbrella window we construct a histogram consisting of $b$ bins. Note that here the bins of the 
histograms do not have to be global, like in WHAM, but each umbrella window is associated with its own set of bins. We then 
calculate the unbiased free energies associated (approximately) with the bin centres (cf.\ Sec.~\ref{sec:UmbSamp}) 
up to an (unknown)\ additive constant which will vary from window to window. We account for the unknown constants by 
introducing them as new hyper-parameters with an uninformative (i.e.\ flat) prior distribution and then integrate the 
posterior over all possible sets of values. Due to the use of Gaussian probability distributions, this integration step 
can be done analytically. 


\subsection{Unknown constants as model parameters}\label{sec:vague}
Writing $\mathbf{y}$ for the noisy free energy readings obtained, we have to modify the expression for the likelihood, 
Eq.~\ref{eq:GPRlikelihood}, to allow for the unknown constants. There are as many unknowns as there are windows 
and we collect them in the vector $\mathbf{f}_0$.  Given a total of $n$ bins across all windows, let  $H$ be a matrix
describing what window each bin 
belongs to: $H_{\theta i}=1$, if bin $i$ belongs to window $\theta$ and $H_{\theta
i}=0$ otherwise.
Given $\mathbf{f}_0$, we can subtract it from  the observed free energy readings and write down the following likelihood, 
\begin{align}
p(\mathbf{y}|\mathbf{f}(\mathbf{x}),\mathbf{f}_0)&=p(\mathbf{y}-H^T\mathbf{f}_0|\mathbf{f}(\mathbf{x}))\nonumber\\
&=(2\pi)^{-n/2}|\Sigma_y|^{-1/2}\times\nonumber\\
\intertext{$\exp\left[-\frac{1}{2}(\mathbf{y}-H^T\mathbf{f}_0-\mathbf{f}(\mathbf{x}))^T\Sigma_y^{-1}(\mathbf{y}-H^T\mathbf{f}_0-\mathbf{f}(\mathbf{x}))\right]$}\label{eq:betalikelihood}
\end{align}
To obtain the posterior distribution, we assume a Gaussian process prior on $f$ and a flat (uninformative) prior on $
\mathbf{f_0}$ and then integrate over $\mathbf{f_0}$. We defer the details of the derivation to Appendix 
\ref{sec:convolutions}, where we obtain a posterior Gaussian 
process with mean,
\begin{align}
\bar{f}(x^*)&=\mathbf{k}^T(x^*)\tilde K\mathbf{y}\label{eq:flatprior_postmean}\\
\tilde K &= \left(K_y^{-1}-K_y^{-1}H^T[HK_y^{-1}H^T]^{-1}HK_y^{-1}\right)\nonumber
,
\end{align}
and covariance,
\begin{equation}
\cov(f(x_1^*),f(x_2^*))=k(x_1^*,x_2^*)-\mathbf{k}^T(x_1^*)\tilde K\mathbf{k}(x_2^*). \label{eq:flatprior_postvar}
\end{equation}

The predictive mean and variance still have the forms of Eq.~\eqref{eq:GPpostmean}
and \eqref{eq:GPpostvar}, so the computational algorithm given in Sec.~\ref{sec:GPR} is thus easily adapted to 
the present case. The effective inverse covariance matrix, $\tilde K$, however, now shows a greater degree of complexity, accounting for our prior ignorance about 
the constants $\mathbf{f}_0$.
The new term  in Eq~\eqref{eq:flatprior_postvar} is positive, which makes
sense: our ignorance about the free energy 
offsets translates into a broader posterior distribution. In Appendix~\ref{sec:diffs} we show that the approach 
presented in this section is numerically equivalent to using the ratio of the bin counts in 
all but one bin and the remaining bin in each window as input data to the GPR. This latter approach does not need 
the introduction and subsequent elimination of the unknown additive constants, since it only considers free energy 
{\em differences} within each window.


\subsection{Modelling the noise structure in the likelihood}\label{sec:likelihood}
The covariances in the previous sections were composed of two parts. The
prior covariance matrix $K$ results directly from the prior covariance function
and the measurement positions ${x_i}$, but $\Sigma_y$, representing the noise
associated with the input data  needs to be estimated. We note that the GPR
formalism necessitates modelling the 
noise as Gaussian in order to remain analytically tractable.

The noise associated with observations from two different windows will clearly be independent and the noise 
covariance matrix will thus have a block diagonal structure. Within one particular window the bin counts 
will have the covariance structure of a multinomial distribution. If all samples were independent this would simply be
\begin{equation}
\cov(n_{\theta i},n_{\theta j})=N_\theta (\delta_{ij}p_{\theta i}-p_{\theta
i} p_{\theta j}),
\end{equation}
where $p_{\theta i}$ are the bin probabilities in the {\em biased} simulation
in window $\theta$.
To account for the time correlation in our samples, we need to consider the effective number of samples 
$N_{\theta,\mathrm{eff}}$, the number of (hypothetical) independent samples required to reproduce the information content 
of the $N_\theta$ correlated samples used\cite{MonteCarloMethods:1989ti}. This can be estimated, e.g., by a block 
averaging\cite{Flyvbjerg:1989ka} procedure or an analysis of the time auto-correlation function\cite{MonteCarloMethods:1989ti}. Given that each effective sample corresponds to $N_\theta/N_{\theta,\mathrm{eff}}$ correlated samples, we obtain
\begin{equation}
\cov(n_{\theta i},n_{\theta j})=N_{\theta,\mathrm{eff}}\left(\frac{N_\theta}
{N_{\theta,\mathrm{eff}}}\right)^2(\delta_{ij}p_{\theta i}-p_{\theta i} p_{\theta
j}).
\end{equation}

Recall that according to Eq.~\refpar{eq:ydef}, up to constants not subject to noise, the observed bin free energies $y_i$, which form the input data for the 
Gaussian process regression, are obtained from the bin counts $n_{\theta i}$ as
\begin{equation}
y_i=-\frac{1}{\beta}\ln \frac{n_{\theta i}}{N_\theta}.
\end{equation}
Propagating errors to first order, we obtain the following estimator for
the covariance,
\begin{align}\label{eq:absnoise}
\cov(y_i,y_j)&\approx
\frac{\partial y_i}{\partial n_{\theta i}}\cov(n_{\theta i},n_{\theta j})\frac{\partial y_j}{\partial n_{\theta j}}\nonumber\\
&=\frac{1}{N_{\theta,\mathrm{eff}}\beta^2}\left(\delta_{ij}\frac{N_\theta n_{\theta i}}{n_{\theta i} n_{\theta j}}-\frac{n_{\theta i} 
n_{\theta j}}{n_{\theta i} n_{\theta j}}\right)\nonumber\\
&=
\frac{1}{N_{\theta,\mathrm{eff}}\beta^2}\left(\delta_{ij}\frac{N_\theta}{n_{\theta i}}-1\right),
\end{align}
where we have also replaced $p_{\theta i}$ with its estimator $n_{\theta
i}/N_\theta$.


\subsection{A binning algorithm}
In this section we address the question of how to best construct the required histograms within each window. This 
problem is outside the scope of the Bayesian analysis presented above, where we have assumed the histogram as 
given. Generally speaking, fewer, larger bins will reduce the statistical noise, but many smaller bins would give 
more detailed information and minimise systematic errors due to associating
the free energies obtained from the bin counts with the bin center.\cite{WHAMerr} The smoothing properties of our GPR 
procedure alleviate this trade-off to some extent, but cannot remove it completely. In the following paragraph we 
describe a binning algorithm we found to work well for the case of a harmonic umbrella potential. We use it 
throughout the rest of this paper.

To obtain the best possible statistics, we place our bins around the measured distribution mean $\mu$ in each 
window and within a range informed by the sample standard deviation $\sigma$. In our numerical examples we will 
use a range of $[\mu-3\sigma,\mu+3\sigma]$.
Rather than splitting the histogram range into the required number of bins evenly, we found better results could be 
obtained by choosing bins that result in similar bin probabilities. We achieved this by choosing bin edges spaced by 
quantiles of a Gaussian with mean $\mu$ and variance $\sigma^2$. The advantage of choosing variable bin widths 
in this way is illustrated by the case of three bins: Three bins of equal width will result in a very accurate central 
reading and two very noisy readings from the tails. Upon taking differences (cf.\ Appendix \ref{sec:diffs}), this results 
in two even noisier difference readings. Three bins of approximately equal bin probability, on the other hand, will result in a 
somewhat noisier observation for the central bin, but the improvement in the tails will more than make up for this.


\section{Bayesian free energy reconstruction from derivative information}\label{sec:GPUI}
Gaussian process regression is not restricted to function reconstructions from noisy function readings and in this 
section we describe how GPR may be used to reconstruct a free energy surface from a number of noisy 
observations of its gradient.

As mentioned above, for GPR to work analytically, both the prior and the
 likelihood need to be  Gaussian. Therefore
observations that are linear functions of the function to be reconstructed
are all admissible. Since differentiation is a linear operator, taking the derivative of a function modelled by a Gaussian process results in a 
new Gaussian process, obtained simply by applying the same operation to the mean and covariance functions.
\cite{Rasmussen:2005ws,Rasmussen:2003ud} For the covariance between values
of the derivative and values of the function we thus have 
\begin{align}
k_{f,f'}(x_1,x_2)&=\left\langle f(x_1)\frac{\partial}{\partial x_2}f(x_2)\right\rangle\nonumber\\
&=\frac{\partial}{\partial x_2}\langle f(x_1)f(x_2)\rangle=\frac{\partial}{\partial x_2}k(x_1,x_2).\label{eq:kff'}
\end{align}
Similarly, the covariance between two derivative values can be obtained as
\begin{equation}
k_{f',f'}(x_1,x_2)=\frac{\partial^2}{\partial x_1 \partial x_2}k(x_1,x_2).\label{eq:kf'f'}
\end{equation}
The posterior mean function given derivative information is then (cf.\ Eq.~\eqref{eq:GPpostmean}
)
\begin{multline}
\bar{f}(x^*)=\mathbf{k}_{f,f'}^T(x^*,\mathbf{x}')(K_{f',f'}(\mathbf{x}')+\Sigma_{y'})^{-1}\mathbf{y'}
,\label{eq:derpostmean}
\end{multline}
where $\mathbf{y'}$ is the vector of (noisy) derivative data obtained at locations $\mathbf{x}'$ and $\Sigma_{y'}$ the 
associated noise covariance matrix. The posterior covariance, analogously to Eq.~\refpar{eq:GPpostvar}, is
\begin{multline}
\cov(f(x_1^*),f(x_2^*))=k(x_1^*,x_2^*)-
\mathbf{k}_{f,f'}^T(x_1^*,\mathbf{x}')\cdot\\
\cdot(K_{f',f'}(\mathbf{x}')+\Sigma_{y'})^{-1}\mathbf{k}_{f,f'}(x_2^*,\mathbf{x}'),
\label{eq:derpostvar}
\end{multline}
  
Just as in the case of function observations (cf.\ Sec.~\ref{sec:GPR}), we can again view the posterior mean as a 
linear combination of kernel functions centred on the data:
\begin{equation}\label{eq:GPRansatz'}
\bar{f}(x^*)=\sum_{i=1}^n b'_i k_{f,f'}(x^*,x'_i),
\end{equation}
where $\mathbf{b}'=(K_{f',f'}(\mathbf{x}')+\Sigma_{y'})^{-1}\mathbf{y'}$. Note, however, that the kernel 
functions are now different ($k_{f,f'}$ rather than the original covariance function $k$), reflecting the different nature of the 
observed data. The example of the SE covariance function, Eq.~\eqref{eq:SE}, illustrates this point clearly: when 
learning from function values, the reconstruction, Eq.~\eqref{eq:GPRansatz}, is a sum of Gaussians centred on the data 
points. When learning from derivatives, the reconstruction is a sum of differentiated Gaussians,
\begin{multline}
k_{f,f'}(x^*,x'_i)=\frac{\partial}{\partial x'_i} k(x^*,x'_i)
=\frac{x^*-x'_i}{l^2}\sigma_f^2\times\\
\exp\left(-\frac{1}{2}\frac{(x^*-x'_i)^2}{l^2}\right).
\end{multline}
At the centres, $x^*=x'_i$, these functions are linear and evaluate to zero, which makes sense because the derivative data provide 
no direct information about absolute function values, only about linear changes.

We apply the technique to umbrella sampling by obtaining free energy derivatives via Eq.~\eqref{eq:UIder}, but we 
stress that any scheme, such as thermodynamic integration\cite{kirkwood,Sprik:1998ud} or ABF\cite{Darve:2001ce}, 
which reconstructs the free energy from its gradients can benefit from GPR.

In addition to estimating the free energy derivatives in each window, we also need to provide estimates for the noise 
in these observations. Just as in Sec.~\ref{sec:likelihood}, we 
again work with the assumption of Gaussian noise, and since the mean force
in each window will be calculated from independent simulations, the noise
covariance  $
\Sigma_{y'}$  is diagonal. We can therefore simply estimate its elements from the variances, $
(\sigma^b_{\theta})^2$, of the biased distributions. We obtain
\begin{equation}
\var\left(y'_{\theta}\right)=\kappa_{\theta}^2\frac{(\sigma^b_{\theta})^2}
{N_{\theta,\mathrm{eff}}},
\label{eq:grad_var_est}
\end{equation}
where $y'_{\theta}$ are the free energy derivatives estimated using Eq.~\eqref{eq:UIder} and $N_{\theta,\mathrm{eff}}$ is 
again the effective number of samples (cf.\ Sec.~\ref{sec:likelihood}).

Modifying the algorithm in Sec.~\ref{sec:GPR} is now straightforward, using
the covariance in \eqref{eq:kf'f'} when precomputing the $\mathbf{b}'$ array and the covariance in \eqref{eq:kff'} when calculating the predicted mean. It is possible to extend the above formalism to incorporate second derivative information estimated from the observed 
sample variances, as suggested by K\"astner and Thiel\cite{UI1}. Since we did not find this to  
improve the quality of our reconstructions, probably because the noise associated with such readings is simply too 
large, we do not report any such results here.


\subsection{Function reconstruction using gradients in several dimensions:\
LSRBF vs. GPR}\label{sec:2+Dtheory}

We now consider the relationship between GPR as shown in the previous section
and a least squares fit using
radial basis functions (LSRBF), in the multi-dimensional case. When reconstructing the free energy surface using a least squares
fit, it is written as a linear combination of
basis functions centred on the data points,
\begin{equation}
f(\mathbf{x}^*)=\sum_{i=1}^n a_i k(\mathbf{x}^*,\mathbf{x}'_i),
\label{eq:RBFansatz}
\end{equation}
where $\{\mathbf{x}_i\}$ are the set of data locations (each is represented
by a vector because we work in more than one dimension)\ and the coefficients, $a_i$, are  obtained by minimising the following error function, given by the sum of the squared deviations of the reconstructed gradients from the observed gradients,
\begin{equation}
E(\mathbf{a})=\sum_{i=1}^n\left\vert\sum_{j=1}^n a_j \boldsymbol{\nabla}_{\mathbf{x}'_i}k(\mathbf{x}'_i,\mathbf{x}'_j)
-\mathbf{y}'_i\right\vert^2,\label{eq:MSerr}
\end{equation}
where $\mathbf{y}'_i$ are  the observed (noisy) free energy gradients.
Like all least squares problems, this minimisation can be reduced
to a linear algebraic system with the following
solution:\cite{Maragliano:2008hw}
\begin{equation}\label{eq:RBcoeffs}
\mathbf{a}=(\boldsymbol{\nabla}\mathbf{k}^T\boldsymbol{\nabla}\mathbf{k})^{-1}\boldsymbol{\nabla}\mathbf{k}^T\mathbf{y'},
\end{equation}
where $\mathbf{y}'$ is the vector obtained by concatenating all the
measured gradients (such that $y'_{iD+\alpha}$ is the $\alpha^{\textrm{th}}$
element of $\mathbf{y}'_i$, where $D$ is the number of dimensions)
and $\boldsymbol{\nabla}\mathbf{k}$ is a $nD\times n$ matrix
containing the gradients of all basis functions at all centres such
that $(\boldsymbol{\nabla}\mathbf{k})_{iD+\alpha,j}$ is the
$\alpha^{\mathrm{th}}$ element of
$\boldsymbol{\nabla}_{\mathbf{x}'_i}k(\mathbf{x}'_i,\mathbf{x}'_j)$.
Maragliano and Vanden-Eijnden further suggest optimising the length
scale (hyper-) parameter, $l$, by minimising the residual
$E(\mathbf{a})$.\cite{Maragliano:2008hw} We show below that this can lead to inferior results
compared to setting a length-scale informed by our {\em a priori}
knowledge about the system.

A comparison of  equations \eqref{eq:GPRansatz'} and \eqref{eq:RBFansatz} shows
that LSRBF and GPR use  similar ansatzes, but differ  in the following important respects: the
nature and number of the basis functions, and the way the coefficients
are calculated. The nature of the basis functions was  addressed in the previous section. The difference in number of basis functions
is unique to the multi-dimensional case: GPR uses one kernel
function in the expansion for every component of the gradient data,
rather than just one for each data point. The multi-dimensional analogue of Eq.~\eqref{eq:GPRansatz'} is thus
\begin{equation}
\bar{f}(\mathbf{x}^*)=\sum_{i=1}^n\sum_{\alpha=1}^D b_{i,\alpha}k_{f,f'_\alpha}(\mathbf{x^*},\mathbf{x}'_i),
\end{equation}
and the $nD\times nD$ covariance matrix
is  given by
\begin{equation}
[K_{f,f'}]_{iD+\alpha,jD+\beta}=\frac{\partial^2}{\partial x'_{i,\alpha}\partial x'_{j,\beta}} k(\mathbf{x}'_i,\mathbf{x}'_j).
\end{equation}

The greater flexibility of GPR afforded by the
increased number of basis functions leads to greater variational
freedom, but the regularisation due to the Bayesian prior mitigates
the risk of overfitting.
While the coefficients in the LSRBF approach are calculated to achieve the best possible fit of the data and therefore risk overfitting, GPR consistently
takes account both of the noise associated with the input data through $\Sigma_{y'}$ and of the expected similarity of different data points through $K_{f',f'}$.
 Both approaches have a computational complexity that
scales similarly with the number of data points, requiring the
construction of a matrix involving gradients of the kernel, and the
inversion of the matrix or the solution of a corresponding linear
system.


\subsection{Using Derivative Information together with Histograms}\label{sec:hybrid}
An advantage of working within a Bayesian framework is that different kinds of information can be 
brought together in a consistent manner. We can thus combine the approaches described in Sections 
\ref{sec:BRUSH} and \ref{sec:GPUI} to obtain a method which incorporates both the mean forces estimated in each 
window and the free energy estimates obtained from histograms in each window.

To achieve this we simply construct an extended covariance matrix between all function values (or function 
differences, cf.\ Appendix~\ref{sec:diffs}) at $\mathbf{x}$ and all derivatives at $\mathbf{x}'$:
\begin{equation}
K_{f\oplus f'}=
\begin{pmatrix}
K(\mathbf{x}) & K_{f,f'}(\mathbf{x},\mathbf{x}')\\
K_{f,f'}^T(\mathbf{x},\mathbf{x}') & K_{f',f'}(\mathbf{x}')
\end{pmatrix}.
\end{equation}
In addition to the covariance matrices considered earlier, it also contains a block, $K_{f,f'}(\mathbf{x},\mathbf{x}')$, 
with the covariances between function and derivative values; this block has elements $[K_{f,f'}]_{ij}=k_{f,f'}(x_i,x'_j)$, 
where $k_{f,f'}$ was defined in Sec.~\ref{sec:GPUI}, Eq.~\eqref{eq:kff'}.

We then make the approximation that the noise associated with the data obtained from our histograms is not 
correlated with the noise in the derivative readings. A combined noise matrix is then simply obtained by forming a block diagonal 
matrix, $\Sigma_{y\oplus y'}=\Sigma_y\oplus \Sigma_{y'}$, out of the original noise matrices.
We also combine the data vectors $\mathbf{y}$ and $\mathbf{y}'$ of the two methods into a new, longer data vector, 
$\mathbf{y}\oplus\mathbf{y}'$ and the covariance vectors $\mathbf{k}(x,\mathbf{x})$ and $\mathbf{k}_{f,f'}(x,
\mathbf{x}')$ 
into the vector $\mathbf{k}_{f,f\oplus f'}(x)=\mathbf{k}(x,\mathbf{x})\oplus\mathbf{k}_{f,f'}(x,\mathbf{x}')$.
Finally we also need to pad the $H$-matrix in these equations with a number
of columns of zero-vectors equal to the 
number of derivative observations (to which
the undetermined constants $\mathbf{f}_0$ do not apply). The 
difference formalism of Appendix~\ref{sec:diffs} does, of course, again provide
an alternative.

We obtain a posterior by substituting these combined vectors and matrices for their counterparts in Eqs. 
\eqref{eq:flatprior_postmean} and \eqref{eq:flatprior_postvar}, i.e.\ $K_{f\oplus f'}+\Sigma_{y\oplus y'}$ for $K_y$, $
\mathbf{k}_{f,f\oplus f'}(x)$ for $\mathbf{k}(x)$ and $\mathbf{y}\oplus \mathbf{y}'$ for $\mathbf{y}$.

We have thus described three variants of the GPR method to reconstruct  free energy profiles. Where we need to 
distinguish them, we shall refer to the methods, respectively, as GPR(h) for the purely histogram based approach of 
Sec.~\ref{sec:BRUSH}, GPR(d) for the approach based on derivatives (Sec.~\ref{sec:GPUI}) and GPR(h+d) for the 
hybrid approach. 

\section{Error estimates}\label{sec:errorbars}

Gaussian process regression  provides an estimate of the variance associated 
with the reconstruction. Care does, however, need to be exercised when this is interpreted in the present context. 
While taking the mean of the  prior to be zero  will result in a reconstructed profile that integrates to zero 
over its range (see below), we stress that the absolute values of the reconstructed free energies are 
meaningful only up to a global additive constant. The diagonals of the covariances calculated using Eq.~
\eqref{eq:flatprior_postvar} or \eqref{eq:derpostvar} must therefore not be interpreted as error bars on absolute free 
energy values. We can, however, use these equations legitimately to estimate the variance associated with free 
energy differences as
\begin{multline}
\var[f(x_1^*)-f(x_2^*)]=\var[f(x_1^*)]+\var[f(x_2^*)]-\\
-2\cov[f(x_1^*),f(x_2^*)].
\end{multline}

Alternatively, we can reinterpret our reconstructed free energy values as 
differences from the global average (zero)\ value of the reconstruction and use the variance on these differences to obtain error 
bars. 
We  distinguish two cases: \begin{enumerate}
\item
 a translationally invariant, periodic covariance function, such as the one given in Eq.~\eqref{eq:periodiccov},  
$k_{2\pi}(x_1,x_2)=\tilde{k}_{2\pi}(x_1-x_2)=\tilde{k}_{2\pi}(x_1-x_2+2m\pi)$, where $m$ is an integer,
\item
 a translationally invariant covariance function which tends to 0 as $x_1-x_2$ tends to $\pm\infty$, such as the SE 
covariance function of Eq.~\eqref{eq:SE},
\end{enumerate}
and show in Appendix~\ref{sec:errormaths} that suitable error estimates are provided
by the variances
\begin{equation}
\var[f(x^*)]-\frac{\bar{k}_{2\pi}}{2\pi},
\end{equation}
and
\begin{equation}
\var[f(x^*)],
\end{equation}
in the respective cases, where 
\begin{equation}
\bar{k}_{2\pi}=\int_0^{2\pi}\tilde{k}_{2\pi}(\tau)d\tau.
\end{equation}


\section{Performance}\label{sec:performance}
We now compare the performance of our GPR reconstruction to those
of WHAM and two variants of Umbrella Integration in one dimension, and to
LSRBF and vFEP in two and four dimensions.
The first UI variant is K\"astner and Thiel's \cite{UI1} original suggestion: the second derivative of the free energy is 
estimated in each window  from the observed variance in the collective coordinate, $(\sigma^b_{\theta})^2$, as
\begin{equation}\label{eq:UIder2}
\frac{\partial^2 F_{\theta}(u)}{\partial u^2}\approx\frac{1}{\beta(\sigma^b_{\theta})^2}-\kappa_{\theta}.
\end{equation}
Like the estimate for the first derivative, Eq.~\eqref{eq:UIder}, this follows from the assumption that the free energy is 
locally (i.e.\ within each window) well approximated by a second order expansion.\cite{UI1}
The second derivatives are then used to linearly extrapolate the first derivatives in the vicinity of the observed 
means.
In the second variant of UI we do not estimate second derivatives from the data, but instead interpolate the mean 
forces from Eq.~\eqref{eq:UIder} using a cubic spline before integrating to yield the free energy profile.


\subsection{Test systems and software}
We shall now  describe our test system, before explaining in more detail how we compare the performance of the 
various free energy reconstruction methods. The latter is not straightforward, because the 
performance of all methods is dependent on the choice of a set of parameters (number of windows, bias 
strength, number of bins), which we cannot expect to be optimal for all methods at once. 
In order to compare the methods in the most fair way, we want each to exhibit its best possible
performance for a given total computational cost.  It may be argued that in any particular application the
optimal parameter settings are not known in advance, but we expect that in practice experience with each
method and system guides the users' choice of parameters {\em a priori}, with only a small amount of 
system-specific tuning. 

We used a model of alanine dipeptide (N-acetyl-alanine-N'-methylamide; Ace-Ala-Nme) to test our methods in one and two dimensions. 
Variants of this system have widely been used in the literature for similar purposes. \cite{Kastner:2009ey, 
Maragliano:2008hw,Bonomi:2009gh} The molecule is modelled with the CHARMM22\cite{MacKerell:1998tp} force 
field in the gas-phase  at a temperature  of 300 K in the NVT ensemble, enforced by a Langevin thermostat
\cite{Schneider:1978ge} with a friction parameter of 100 fs. The equations of motion were integrated with a time step of 0.5 fs.
The calculations were carried out with the LAMMPS\cite{Plimpton:1995fc} package augmented by the PLUMED
\cite{Bonomi:2009gh} library. For 
the WHAM reconstructions reported in this paper, Alan Grossfield's code\cite{WHAM:code} has been used.
The vFEP reconstructions were done with the publicly available software.\cite{vFEP_software}

Two slow degrees of freedom, the two backbone dihedral angles $\Phi$ (C-N-C$_{\alpha}$-C) and $\Psi$ (N-C$_
\alpha$-C-N), are present in this system. 
A one-dimensional system was obtained by adding a harmonic restraining potential in $\Phi$ to the system, centred 
at $\Phi=-2.0$ with a force constant of 100~kcal/mol,  while using $\Psi$ as the collective variable of interest. This 
choice yields a system where all complementary degrees of freedom are fast, so that problems with metastability are 
avoided. This may seem like a highly artificial one dimensional system from a chemist's point of view, but it serves 
the purpose of testing a variety of free energy reconstruction methods. To obtain a converged reference free energy profile  to which all methods are to be 
compared, we ran an umbrella sampling calculation consisting of 50 evenly
spaced windows with a bias 
strength of $\kappa =100$~kcal/mol. Within each window a trajectory was propagated
for 50~ns from which samples were 
taken every 500 fs. A reference curve was constructed from this data  such
that the  root
mean square deviation between using different methods to construct the reference
was less than 10$^{-2}$~$k_\mathrm{B} T$, more than precise enough for our purposes.

The collective variables in the two 
dimensional test case are the two backbone dihedral angles. We collected data on an
evenly spaced grid of $48\times48$ centres with bias strengths
of $\kappa=$~50, 100, 200, and 400~kcal/mol in both collective variables and 500-ps-long
trajectories at each center.  As discussed in more detail below, all GPR and LSRBF
results use $\kappa=100$~kcal/mol, while both $\kappa=50$~kcal/mol and $\kappa=200$~kcal/mol
are used for vFEP.  All two-dimensional free
energy reconstructions are performed using  subsets of this dataset, unless
specifically stated otherwise. 

To test the reconstruction in four-dimensions
we used the same simulation parameters, but replaced the dipeptide with alanine tripeptide (blocked dialanine,
Ace-Ala-Ala-Nme) and used its four backbone dihedral angles as
collective variables. We used two interlocking grids of
$6^4$ evenly spaced umbrella centres each (i.e.\ the second grid
is obtained by translating the first by half a unit cell in all
four directions) as well as a denser grid of $9^4$ evenly spaced
centres, all with a bias strength of $\kappa=400$~kcal/mol
and running trajectories of 500~ps at each center.

The sample
noise increases with increasing umbrella strength (due to equipartition),
but a stiffer umbrella also reduces the non-uniformity of the grid
of mean sample positions, which are displaced from the umbrella
centres by amounts proportional to the local gradients and inversely
proportional to the umbrella strength.  We find that in the four-dimensional
case,
when data is intrinsically more scarce, the stiffer bias potential gives better results than the softer one
used in two dimensions.

For GPR we  used a prior function variance
of  $\sigma_f^2=10.0$~kcal$^2$/mol$^2$ in the one dimensional case but  $\sigma_f^2=20$~kcal$^2$/mol$^2$ for the two and four dimensional cases since there a larger range is expected.
The length scale parameter was $l=\pi/3$ unless stated otherwise.

In the GPR and LSRBF approaches the periodicity of 
the reconstructed free energy profile is guaranteed through the use of a periodic covariance
 function (Eq.~\eqref{eq:periodiccov}). 
In order to perform a fair comparison, we therefore must also enforce periodicity
when testing other interpolation methods (UI in particular). We do 
this by linearly subtracting fractions of the integral over a full rotation,
\begin{equation}\label{eq:periodize}
F(\Psi)=\int_0^{\Psi} F^\prime_{\mathrm{UI}}(\tilde{\Psi})d\tilde{\Psi}-
\frac{\Psi}{2\pi}\int_0^{2\pi} F^\prime_{\mathrm{UI}}(\tilde{\Psi})d\tilde{\Psi},
\end{equation}
where $\Psi$ is a dihedral angle, and $F_{\mathrm{UI}}^\prime$ is the interpolated
mean force from the umbrella 
simulations. This simple ad-hoc adjustment  may be considered to be an extension
of the original UI to periodic 
systems.\cite{Kastner:2009ey}

GPR needs an estimate of the noise associated with
each measurement, but especially for the shortest
trajectories in two and more dimensions, the estimate of Eq.~\ref{eq:grad_var_est} evaluated
for a single gradient may involve only a handful of samples, and therefore 
be inaccurate.  Instead, for each trajectory length up to 5~ps, we use {\em
all}
the
samples to estimate the error in the individual gradient sample, and use that 
value in the GPR for all  gradient components.


\subsection{Performance in one dimension}

In order to select the best parameter settings for each reconstruction method
we  performed five sets of umbrella sampling calculations, each with 24 evenly distributed windows but 
differing bias strengths. Within each window we ran a 5~ns-long trajectory, recording samples every 50 fs. The 
data was then used to create repeated free energy reconstructions using a range of 
parameter choices and for different amounts of total trajectory used. The amount of data used was varied by factors of 10 
from 12~ns per reconstruction (1/10 of the generated data) down to just 12~ps per reconstruction (1/$10^4$ of the 
available data). For each of these data amounts, the following parameters were varied and  reconstructions 
produced for all cases:
\begin{itemize}
\item
Umbrella strengths, $\kappa$: 5, 15, 25, 50 and 100~kcal/mol
\item
Number of windows: 8, 12, and 24, while keeping the total data used constant by varying the trajectory length 
per window (we also checked using fewer windows, but these reconstructions were very clearly inferior)
\item
For WHAM and histogram-based GPR reconstructions, the number of bins: 20, 40 and 80 bins for WHAM (we 
deemed a reconstruction with fewer than 20 bins as too coarse to be meaningful, while significantly more than 80 
bins resulted in very noisy reconstructions) and 2-10 bins per window for GPR(h) 
and GPR(h+d)
\end{itemize}

\begin{figure}[htbp]
\centering
\includegraphics{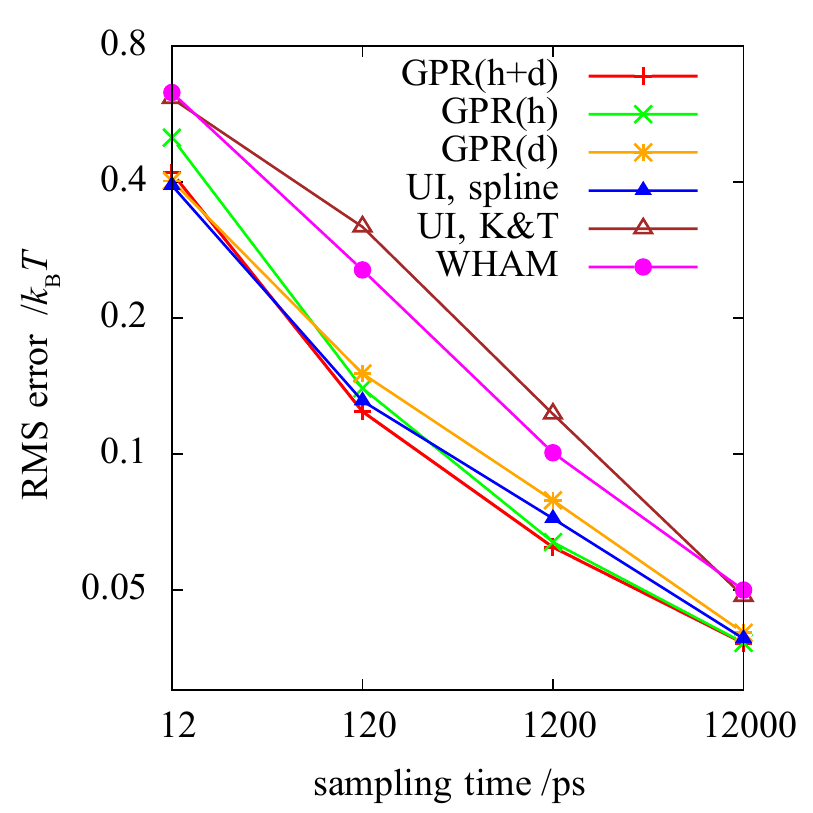}
\caption{Comparison of the  average root mean squared errors of a number of free energy reconstruction 
methods as a function of total sampling time using optimal parameter
settings for each method and sampling time.\label{fig:masterplot}}
\end{figure}

\begin{figure*}[htbp]
\centering
\includegraphics[width=15cm]{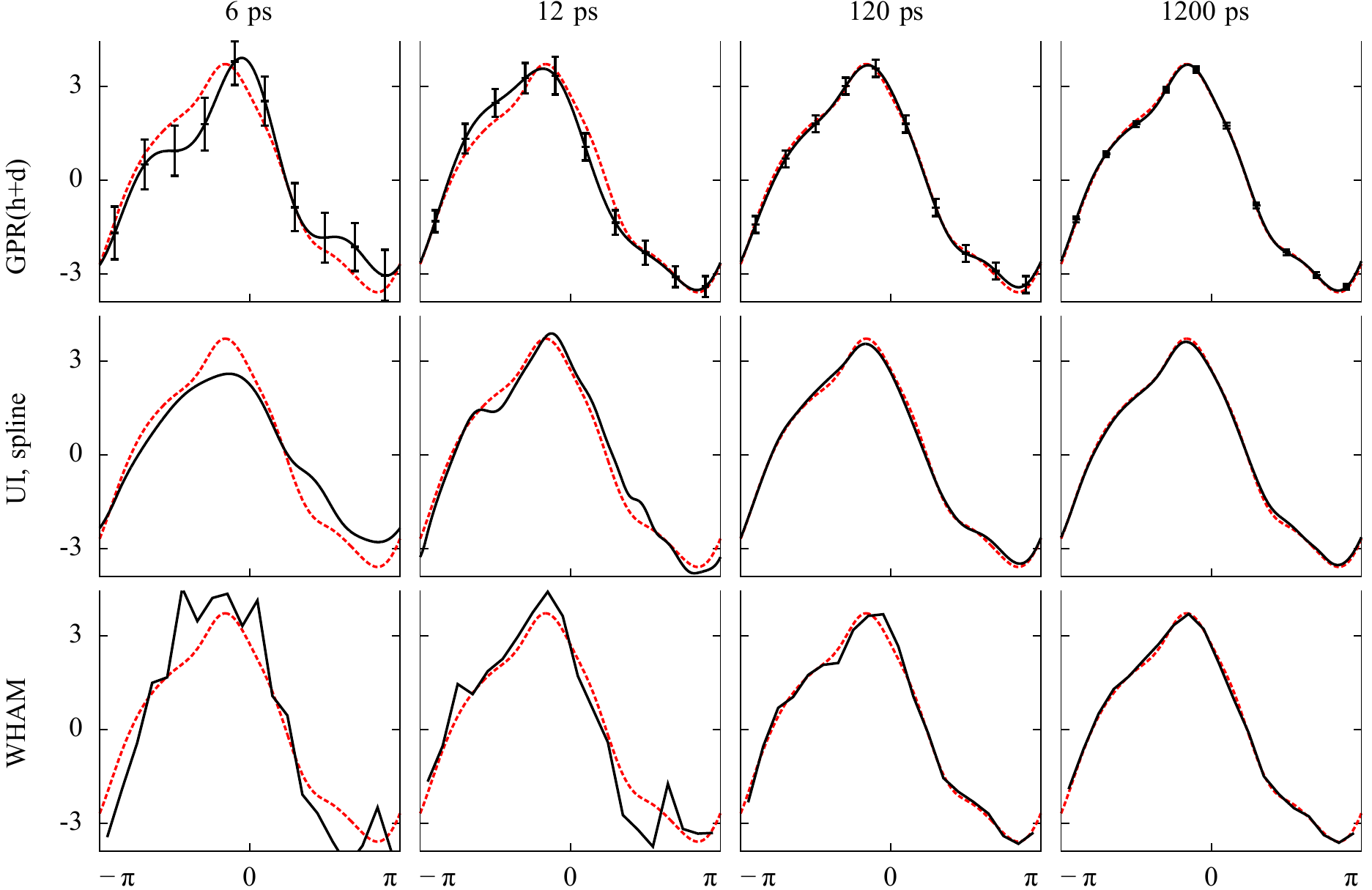}
\caption{Representative free energy reconstructions (in units of $k_{\mathrm{B}}T$) using different methods for total sampling times of 6, 12, 120 and 1200~ps. 
The dashed red curve is an accurate reference. The error bars for GPR(h+d) represent two standard deviations 
calculated using Eq.~\eqref{eq:errorbars}.}
\label{fig:repplots}
\end{figure*}

For each choice of method, total amount of data, and other parameters, 100 reconstructions were carried out using 
different sections of the full 5~ns trajectories. The root mean squared (RMS) error of these 100 reconstructions with 
respect to the reference curve was then calculated (in the case of WHAM at the bin centres only). Because  all of the 
reconstructions are determined up to an additive constant only, this was chosen in each case to obtain a reconstruction with a global average value
of zero. The optimal parameter settings for each method are summarised in Table~\ref{tab:optimalparas}.
These best-case results were then used to compare the performance of the various methods with each other, as shown in 
Figure~\ref{fig:masterplot}.  Figure~\ref{fig:repplots} shows representative reconstructions for some of the methods using 
different amounts of input data.  

\begin{table}[htbp]
\caption{Optimised parameter choices for a number of free energy reconstruction methods given different amounts of 
total sampling time.\label{tab:optimalparas}}
\begin{tabular}{c|c|cccc}
\hline  &&12~ns&1.2~ns&120~ps&12~ps\\
\hline 
        GPR&$\kappa$&100&25&15&15\\
        (h+d)& win \#&12&24&24&8\\
        & bin \#&5&2&2&2\\
        \hline
        GPR&$\kappa$&25&15&15&15\\
        (h)& win \#&24&24&24&8\\
        & bin \#&2&2&2&2\\
        \hline
        GPR&$\kappa$&100&25&15&15\\
        (d)& win \#&24&24&24&12\\
        \hline
        UI&$\kappa$&100&25&15&25\\
        spline&win \#&24&24&24&12\\
        \hline
        UI&$\kappa$&15&15&15&15\\
        K\&T& win \#&12&24&24&24\\
        \hline
        WHAM&$\kappa$&15&15&15&15\\
        & win \#&24&24&24&24\\
        & bin \#&20&20&20&20\\
\hline
\end{tabular}
\end{table}

The traditional method rivalling the GPR reconstructions most closely in this comparative study is the spline based UI. We note 
that regression using one-dimensional cubic splines as basis functions can be thought of as a special case of Gaussian processes\cite{Rasmussen:2005ws},
so a similar performance should perhaps not be very surprising. UI does not, however, allow for 
consistent treatment of measurement noise in the data. While in one dimension this does not seem to have a 
significant detrimental effect on the quality of the reconstructions as measured by the RMS error, the corresponding 
reconstructions in Figure~\ref{fig:repplots} do seem to be less smooth than their GPR counterparts, with the example corresponding to the 12 ps 
trajectory even showing false local minima.  
The widely used WHAM is seen to produce very noisy reconstructions, which are improved upon by our approaches 
both visually and measurably in the RMS error.

Figure~\ref{fig:masterplot} also shows that K\"astner and Thiel's\cite{UI1}  interpolation of mean forces---which uses 
measured second derivative information---performs rather less well than the spline interpolation which makes no 
reference to second derivatives. The reason for this is that the  first and second 
derivative information enter the fit with a similar weight (particularly so if the windows are widely spaced), despite 
the fact that the latter typically has a much larger statistical error associated with it. 

It is interesting to note that the optimal  number of bins per window for GPR(h) and GPR(h+d) was two, independent of other parameters. We speculate that this is a result of the better statistics that can be achieved using fewer bins. 
Indeed, the situation has parallels to the estimation of free energies from derivatives. The information from a two-bin 
histogram is a single free energy difference, which is closely related to the information from a single free energy 
derivative, although without the approximation of the distribution mode by its mean (Eq.~\eqref{eq:UIder}).

Finally, we observe that the methods based on mean force estimates from Eq.~\eqref{eq:UIder} (i.e.\ spline based UI 
and GPR(d)) slightly outperform the GPR reconstruction based on histograms only (GPR(h)) in the limit of very short 
sampling trajectories, while the opposite is true in the limit of long sampling. A tentative explanation is that 
in the abundant data regime statistical errors become small enough for systematic errors to dominate. Since Eq.~
\eqref{eq:UIder} is based on a second order expansion of the free energy it introduces such a systematic error, 
which is apparently larger than the error made by constructing histograms. In the opposite regime statistical errors 
dominate, leading to two adverse effects for GPR(h). First, when using histograms, the data in each window have to 
be divided between at least two bins, whereas all data are used to obtain the estimate Eq.~\eqref{eq:UIder}, thus 
resulting in better statistics. Furthermore, the assumption made in GPR(h) that the noise associated with the bin free 
energies can be approximated as Gaussian (cf.\ Sec.~\ref{sec:likelihood}), will start to break down in the limit of very 
short sampling trajectories. The hybrid GPR(h+d), finally, appears to fulfil our expectations of being the most 
accurate overall.


\subsection{Two dimensions}
\label{sec:two_dimensions}

In two dimensions we compare GPR(d) to the established LSRBF method, and the
recently proposed vFEP approach\cite{Lee_JCTC_2013,Lee_JCTC_2014}. The latter models the underlying free energy with cubic splines and uses the maximum likelihood principle
on the histogram collected in all windows to optimise the spline parameters. It has been shown to significantly outperform
both WHAM and MBAR, which also operate directly on the histogram, in two dimensions.\cite{Lee_JCTC_2014}

In the following, where we report RMS errors with respect to the reference,
we mainly compare {\em gradients} rather than the free energy profiles. This is because we prefer not to favour any of the reconstruction methods by selecting
one to define the reference profile. The gradients, on the other hand, are direct outputs of the umbrella simulations, and can therefore be converged very accurately by running long simulations without the need for any postprocessing.        

\begin{figure}[htbp]
\centering
\includegraphics[width=8cm]{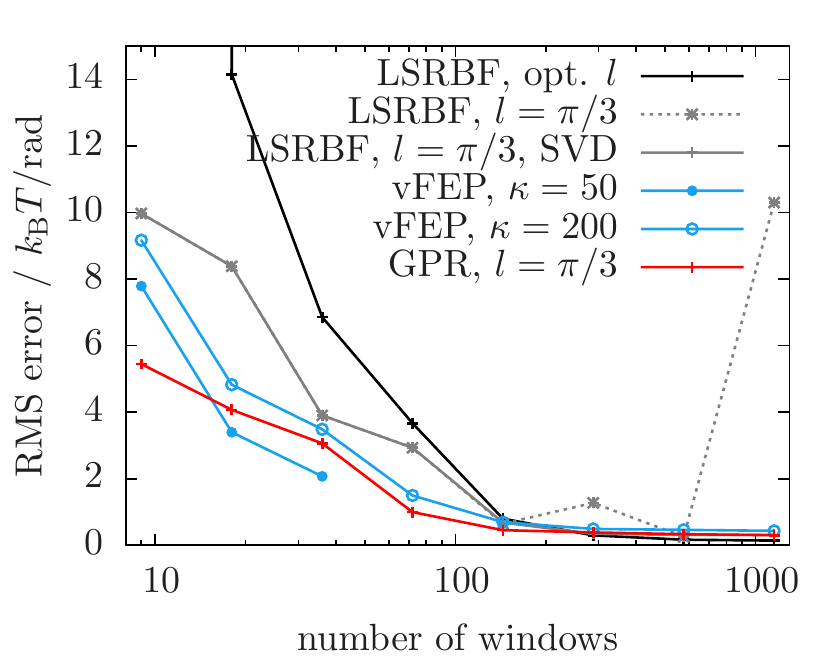}
\caption{RMS deviations of free energy gradients
of alanine dipeptide reconstructed using LSRBF
(with and without singular value decomposition
to control the condition number), vFEP (with two values of $\kappa$), and GPR(d).  \label{fig:2Dders_nwin}}
\end{figure}

We first investigate the performance of the three methods as a function of
grid size,
i.e.\ the number of umbrella windows, 
  using the entire 500~ps trajectory from each window in order to keep
the statistical noise low.  Figure \ref{fig:2Dders_nwin} shows the RMS gradient error in the
reconstruction of the two dimensional free energy surface. The error is given as the mean squared
deviation of the analytical gradients of each reconstruction from
the measurements made on the densest ($48\times48$) grid.  We do
find, as expected, that the condition numbers of the matrices in
the LSRBF reconstructions are often very large indeed, at times
exceeding values of $10^{18}$. This leads to an instability of the
linear solver due to finite precision floating point arithmetic,
which is easily controlled by employing a singular value decomposition
(SVD), for which we use a relative threshold of $10^{-12}$.

We also find that the publicly available vFEP implementation is
very sensitive to the strength of the bias in the sampling. Data
generated using a moderate spring constant of $\kappa=50$~kcal/mol
gave the best results when the grid was sparse, but on the denser
grids we were unable to obtain a sensible reconstruction when
attempting to process data that was generated using 50 and 100~kcal/mol,
and only the data corresponding to $\kappa=200$~kcal/mol was usable.
Since $\kappa$ is a parameter of the umbrella simulations themselves
rather than that of the reconstruction, tuning it during the reconstruction
process is not in general
a practical option, because it would require rerunning the
computationally expensive sampling.  We therefore settle on one
value, $\kappa=200$~kcal/mol, for all vFEP results that follow in
order to be sure that reconstructions can always be made. LSRBF and
GPR(d) are much less sensitive to the umbrella strength. Higher values
of $\kappa$ help to  reduce the systematic
errors associated with the approximation in
Eq.~\eqref{eq:UIder}, but also lead to higher statistical noise and could
potentially induce stiff dynamics that are hard to sample.

\begin{figure*}[hp]
\centering
\includegraphics{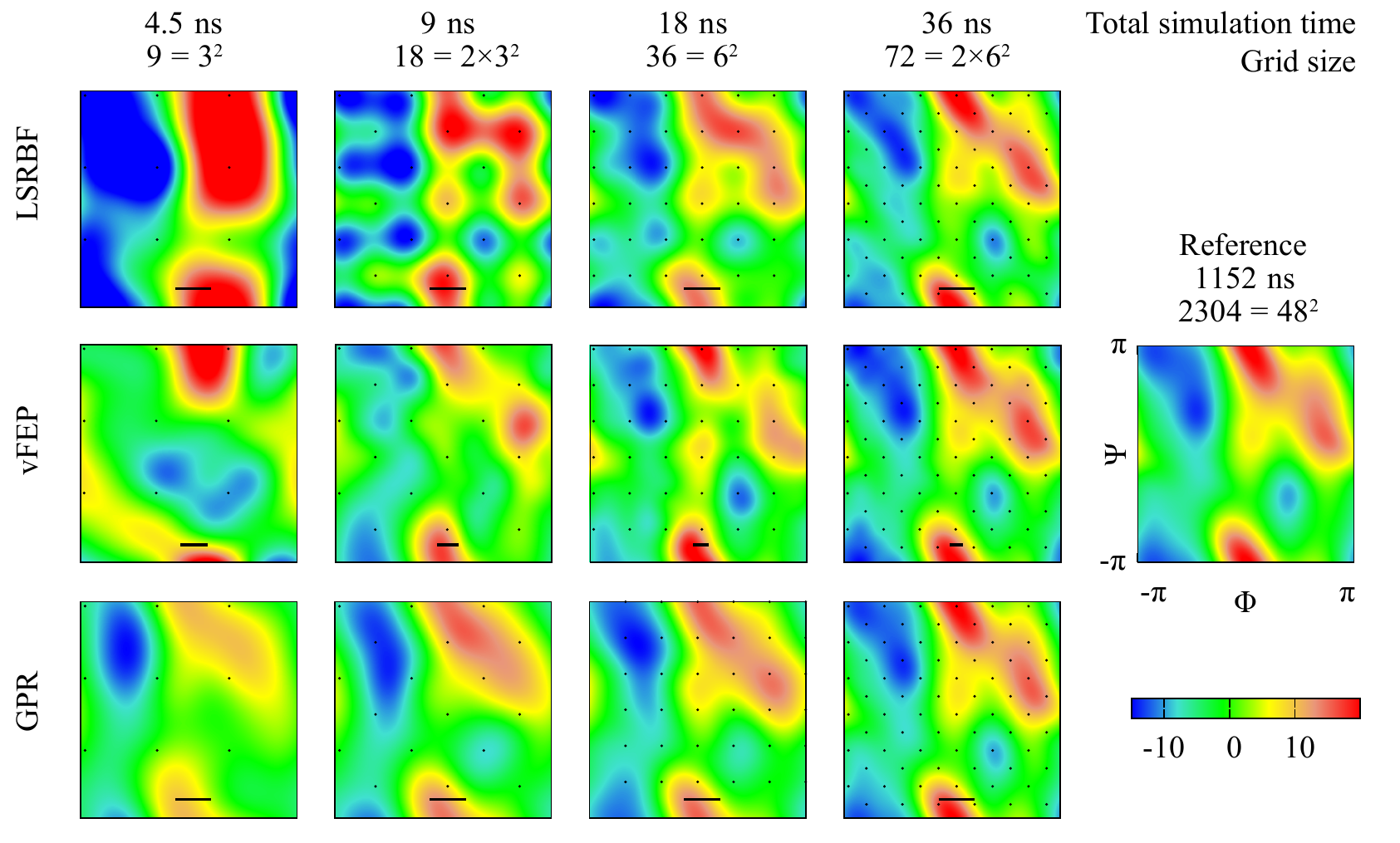}
\caption{LSRBF, vFEP and GPR(d) reconstructions of the free energy surface
(in units of $k_\mathbf{B} T$)
of alanine dipeptide in two dimensions using different numbers of
windows, with a sampling trajectory of 500~ps/center. The LSRBF and GPR(d) reconstructions
use $\kappa=100$~kcal/mol, while vFEP use $\kappa=200$~kcal/mol. The
reconstructions are visually indistinguishable when more than 72
centres are used. 
The black dots mark the location of the bias centres. The black line segments indicate our choice
of length-scale parameter $l=\pi/3$ for GPR(d) and LSRBF, and in case of vFEP they show the spacing between spline points chosen by the vFEP software, corresponding to $n_s=8, 11, 15, 17$, respectively.   \label{fig:2Dvs}}
\end{figure*}

While the differences between the respective RMS 
gradient errors are not dramatic for the denser grids, in the regime of sparse
grids both vFEP and GPR(d) 
outperform LSRBF.  The visualisations of the free energy surface
reconstructions, shown in Figure~\ref{fig:2Dvs}, underscore the advantages
of GPR(d).

\begin{figure}[htbp]
\centering
\includegraphics[width=8cm]{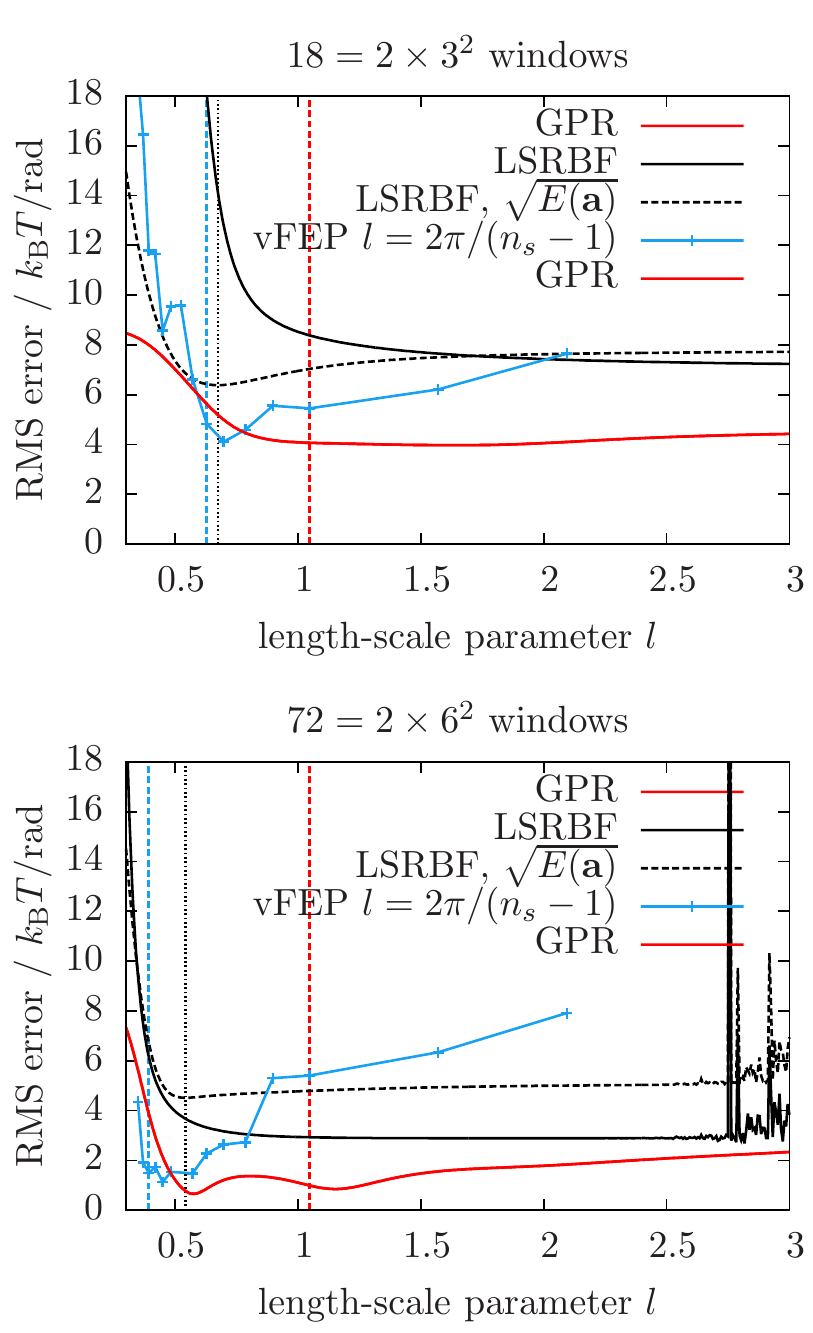}
\caption{Variation of the RMS gradient error as a function of length scale for GPR(d), LSRBF, and vFEP,
using 18 (top) and 72 (bottom) centres.  For GPR(d) and LSRBF the length scale is the hyperparameter $l$,
while for vFEP we use the spacing between spline points.  Also shown is the quantity $\sqrt{E(\mathbf{a})}$ for the 
LSRBF reconstruction, used in Ref.~\citenum{Maragliano:2008hw} to optimise $l$; its minimum value is marked by 
the dotted blue vertical line. The dashed black vertical line marks the value $l=\pi/3$. The dashed blue line corresponds to the automatic choice of spline
point spacing of the vFEP implementation.\label{fig:2Dders_l}}
\end{figure}

 Figure~\ref{fig:2Dders_l} shows the
variation of the RMS gradient error of GPR(d) and LSRBF with the choice
of length scale parameter for two different grid sizes, as well as
the variation of the square root of the quantity $E(\mathbf{a})$,
defined in Eq.~\eqref{eq:MSerr}, which  is used in
Ref.~\citenum{Maragliano:2008hw} to find an ``optimal'' value for $l$
for LSRBF.
Optimising the length-scale hyper-parameter $l$  does not offer any great advantage over our{\em\ a priori}
choice of
 $l=\pi/3$.
Indeed it can  lead to
very unphysical results, particularly in the limit of scarce data in the LSRBF case, because it exacerbates the problem of
overfitting.  It is our view that for chemical and material systems it is 
usually not very difficult to choose a reasonable value of $l$ before the reconstruction is undertaken,
taking note of the  meaning of this hyper-parameter: it is
the expected correlation length of the free energy surface. For smooth
functions in dihedral angles, setting it to any value in the range $[\pi/3,\pi]$ is reasonable, as demonstrated
by the relatively flat curves of the RMS gradient error for $l$
values in this range. We expect this to hold true for all free
energy surfaces in dihedral angles of molecular systems.

We also show  on Figure~\ref{fig:2Dders_l} the performance of vFEP as  the number of spline points $n_s$ is varied, which controls the smoothness of its reconstruction (in order to use the same horizontal scale, we made the identification
$l=2\pi/(n_s-1)$). The error of the reconstruction changes significantly as $n_s$ is varied, in contrast to the relative tolerance of LSRBF and GPR(d) with respect to their length scale parameters. Furthermore, the range of good parameters for the latter methods mostly depends on the properties of the underlying function, whereas the narrow range of $n_s$ for which vFEP gives a good reconstruction varies  with the grid density. The implementation of the vFEP method we used does offer a default  choice for $n_s$, indicated
by the gray dashed line, which seems to work well in the present case in
which there is very little statistical noise in the data.

\begin{figure}[]
\centering
\includegraphics[width=8cm]{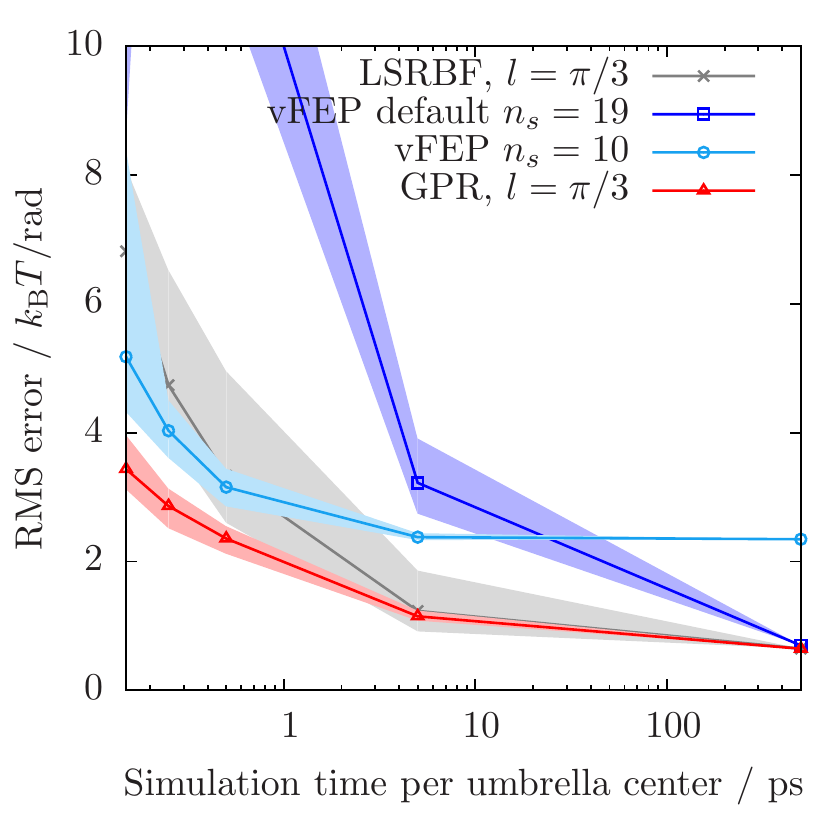}
\caption{Median and
typical range of the RMS deviations in
the reconstructed free energy gradients of alanine dipeptide. Samples
of 50 reconstructions for each trajectory length and method  were used to calculate the median; the shaded area represents
the range between the 5th and the 95th percentile of these samples
(i.e.\ between the third best and third worst sample out of 50). All
reconstructions use a grid of $12\times 12$ windows, while
the sampling time in each window is varied as
shown.\label{fig:GPvRB_noisy}}
\end{figure}

We now explore the relative performance of the  methods in
the regime of high statistical noise. To this end, we reconstructed
the free energy surface from successively shorter trajectories
sampled on a $12 \times 12$ grid of umbrella centres.  As the
trajectory length decreases, the results obtained show an
increasing variance between runs. 
Figure~\ref{fig:GPvRB_noisy} shows
the medians of the RMS reconstruction gradient errors obtained from
50 different reconstructions with the same trajectory length. We also
show an estimate of the breadth of the distribution, by plotting the range
between the 5th and the 95th percentile of the sample of reconstructions.

In this regime of a dense grid and high noise  the relative performance of vFEP and LSRBF are now opposite as compared with
the previous low noise case.  The vFEP method shows much worse degradation than LSRBF as the noise is increased, 
while GPR(d) outperforms both. Furthermore, while the performance of vFEP can be improved in the high noise limit by halving the number of spline points
from the default choice (e.g.\ $n_s=10$ instead of $n_s=19$ for the plotted $12 \times 12$ grid) and forcing a smoother reconstruction, 
this reduction in variational freedom results less accurate reconstructions for 
the intermediate and low noise cases.  

\begin{figure*}[htbp]
\centering
\includegraphics{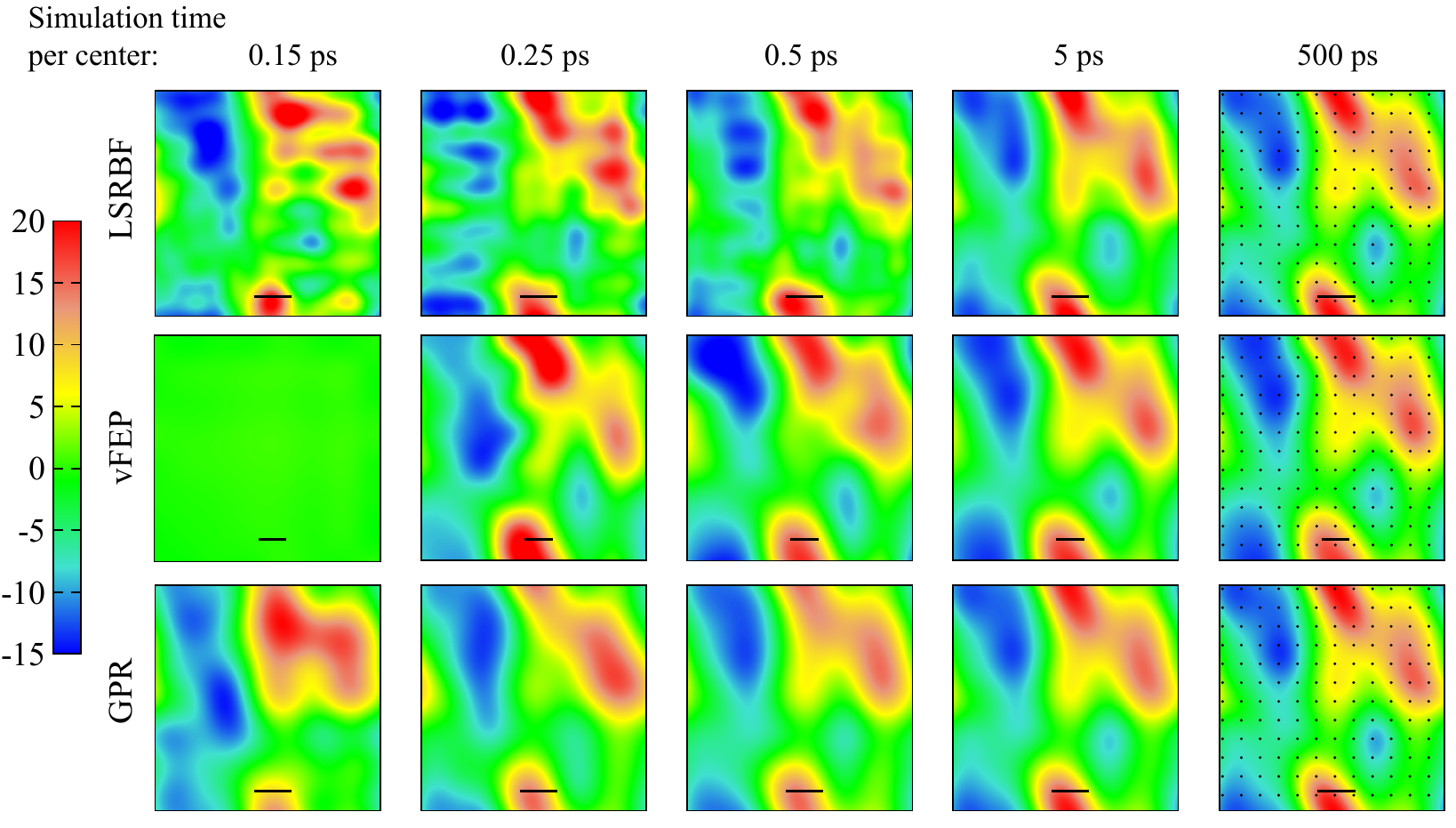} 
\caption{LSRBF, vFEP and GPR(d) reconstructions of the free energy surface
(in units of $k_{\mathrm B} T$) of alanine dipeptide
from data gathered in  a  $12\times 12$ grid of windows over a varying amount
of sampling time. The bias centres
are marked by black dots in the rightmost column, but apply for all panels. We show 
the reconstructions with the third-largest RMS gradient errors from
a sample of 50 (representing the 95th percentile) for each sampling
time. The black line segments indicate the magnitude of the
length-scale parameter $l=\pi/3$ for GPR(d) and LSRBF, and the spacing between the spline points for vFEP corresponding to $n_s=10$ spline points.  \label{fig:2Dvs_noisy}}
\end{figure*}

To give a feel for the kinds of reconstructions obtained, Figure~\ref{fig:2Dvs_noisy}
shows the reconstructions corresponding to  approximately the 95th percentile of the sample
(the third-worst out of 50) for each trajectory length. We  show reconstructions
from the worse end of the distribution as they
better highlight the differences in  resilience to noisy input
data. While the LSRBF and vFEP reconstructions attempt to fit the data
(including the noise) as closely as possible, the GPR(d) reconstructions are
better in taking
the statistical noise into
account, resulting in a greater resilience to it. The flat green ``reconstruction''
of vFEP using data from only 0.15~ps trajectories corresponds to all zeros---there simply was not enough data there for the vFEP procedure to avoid converging to the null result.   

\begin{figure}[]
\centering
\includegraphics[width=8cm]{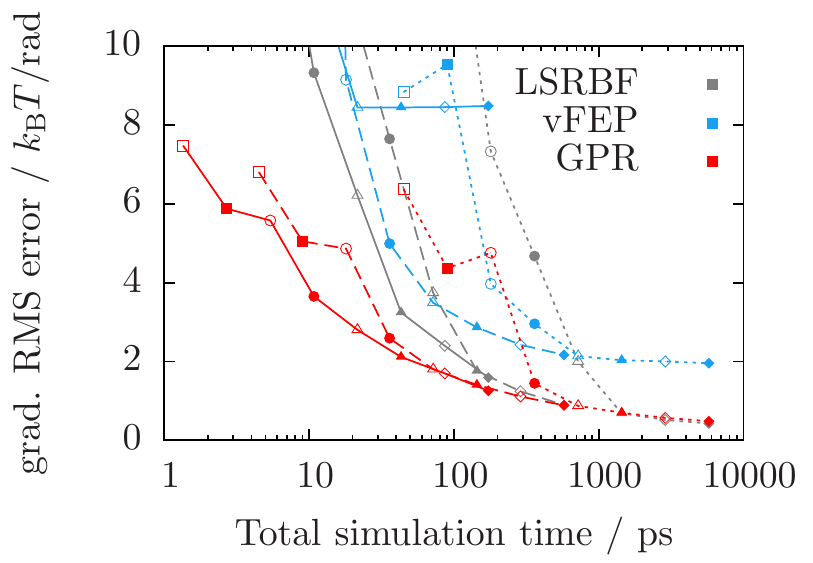}
\includegraphics[width=8cm]{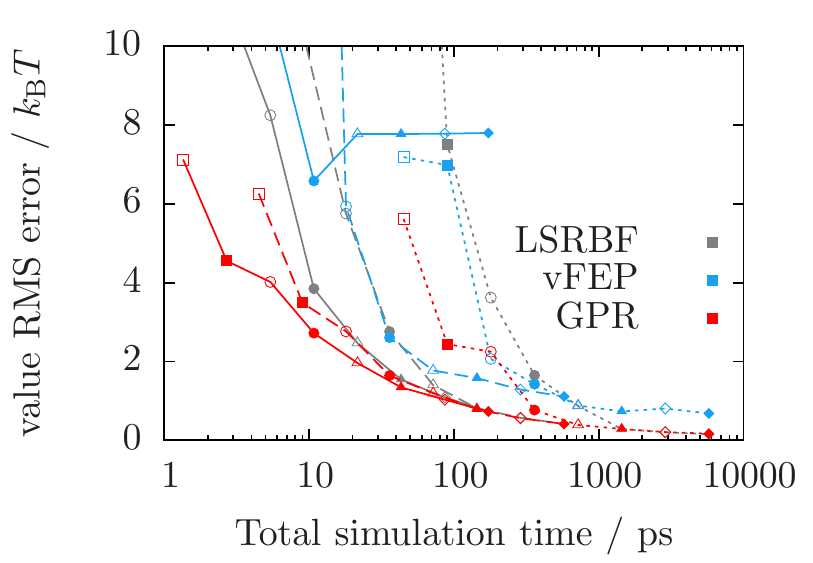}
\caption{RMS error of gradient (top panel) and value (bottom panel) for vFEP (blue), RBF (gray), and GPR(d) (red) as a function of total amount of simulation time, for a variety of grids and trajectory
lengths per grid point, showing the 3rd worst result from a sample of 50
reconstructions.  Line type indicates sampling time per window: 0.15~ps (solid), 0.5~ps (dashed), 5~ps (dotted).
Symbol indicates grid size: $3\times3$ (open square), $3\times3\times2$ (filled square), $6\times6$ (open cicle), $6\times6\times2$ (filled circle), $12\times12$ (open triangle), $12\times12\times2$ (filled triangle), $24\times24$ (open diamond), $24\times24\times2$ (filled diamond).
\label{fig:RMSvswork}}
\end{figure}

So far we have showed reconstructions for various grid sizes at a fixed cost/grid
point and also for varying trajectory length at a fixed grid size. But in practice
one would be interested in the best reconstruction given a fixed total amount
of computational cost, thus making a tradeoff between more grid points
or more samples collected at each point. Figure~\ref{fig:RMSvswork} shows the results combined
for all grid sizes and trajectory lengths (note that the total simulation time on the $x$ axis does
not include any time needed for equilibration at each umbrella position).
In order to attempt a fair comparison, we again halved the number of spline points compared to the default choice  made by vFEP for short sampling times. For all grids and trajectory lengths we show the 3rd worst reconstruction out of 50 samples.

Since it is not necessarily obvious what the error in the free energy
will be as a function of the error in the gradient, here we plot both in two separate panels.  
 The reference profile for the free energy error is the result of a GPR reconstruction
based on all the data, i.e.\ 500~ps per window on the $48 \times 48$ grid.

Irrespective of the method used to do the reconstruction,  it turns
out to be always better to choose a denser grid and shorter
trajectories, which is remarkable given the comparatively poor noise
tolerance of LSRBF and vFEP.  In a more realistic application, where
equilibration time is included, the optimal tradeoff is unclear.
The use of denser grid would seem to require additional equilibration
because each grid point needs to be equilibrated separately.  However,
if each grid point's trajectory is started from an adjacent one, the
smaller distance from one grid point to the next might shorten the
required equilibration time at each point.

\subsection{Four dimensions}
\label{sec:four_dimensions}

\begin{figure}[htbp]
\centering
\includegraphics[width=8cm]{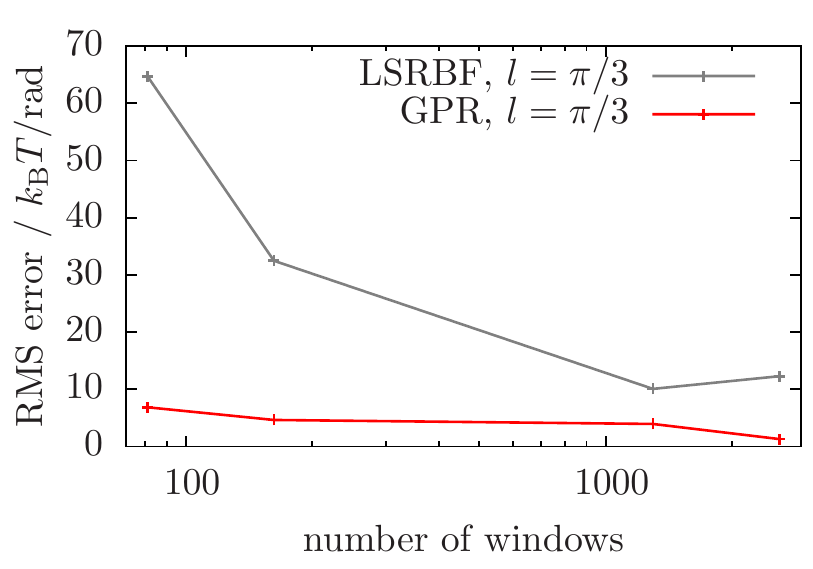}
\caption{RMS deviations of free energy gradients reconstructed using
LSRBF and GPR(d) in four dimensions from subsets of $2\times6^4$
free energy gradient observations, with respect to a reference set
of $9^4$ free energy gradients. \label{fig:4Dders}}
\end{figure}

Finally, we compare the performance of the LSRBF and GPR(d) methods in four
dimensions, where data is necessarily much scarcer (the publicly available
implementation of vFEP does not support more than two dimensions).  
Figure~\ref{fig:4Dders} shows the RMS deviations of free energy
gradients obtained from reconstructions using various subsets of
the data  with respect to a reference  obtained on the densest grid of $9^4$ evenly spaced umbrella centres. GPR(d) outperforms the LSRBF fit, but even using more
than 2000 windows GPR(d) has a remaining gradient error of
about $k_{\mathrm{B}}T/\mathrm{rad}$.  The LSRBF, on the other hand, fails to
provide a reconstruction better than 10~$k_{\mathrm{B}}T/\mathrm{rad}$ even when
using the densest grid.

\begin{figure}[htbp]
\centering
\includegraphics{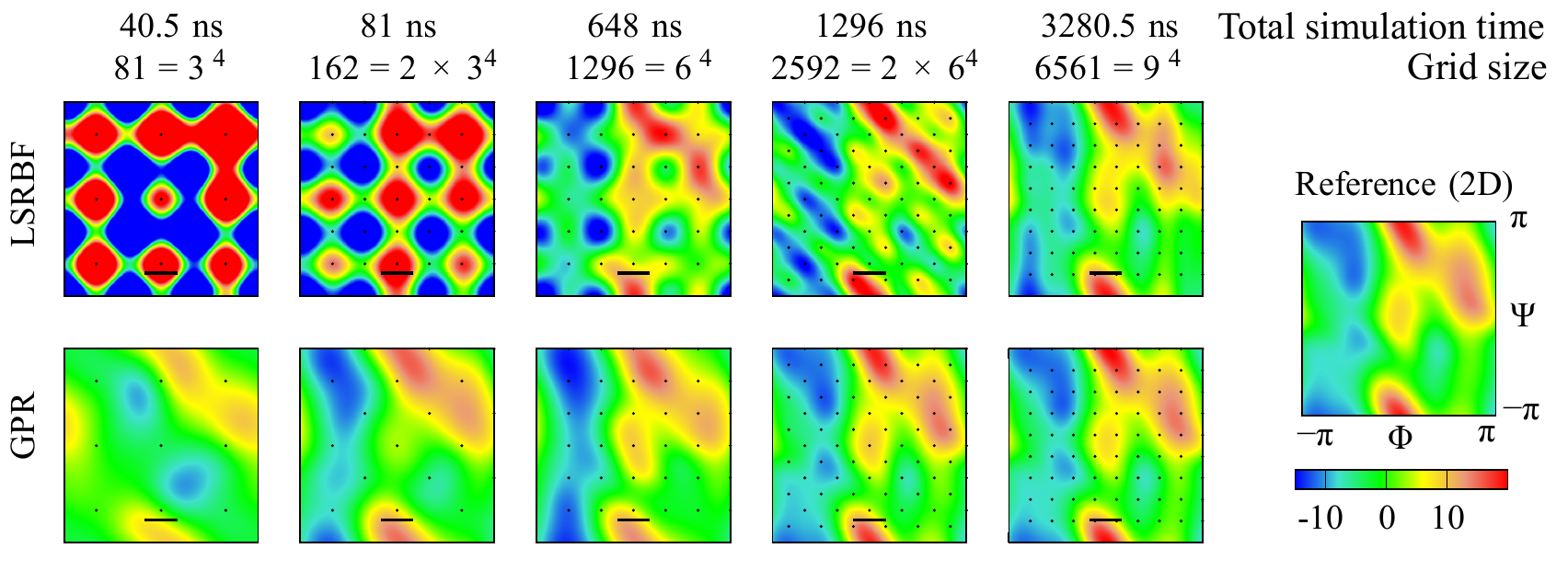}
\caption{2D slices of GPR(d) and LSRBF reconstructions of the 4D free
energy surface (in units of $k_{\mathrm B} T$) of alanine tripeptide
using different numbers of windows, at third and fourth
backbone dihedrals (from the N-terminus) of -2.0 and 2.0, respectively.
$\Phi$ and $\Psi$ refer to the first two backbone dihedrals from
the N-terminus, respectively. The black dots mark the locations of the bias centres projected onto the 2D slice of the reconstruction.
The black line segments indicate the magnitude of length-scale
parameter $l=\pi/3$. 
\label{fig:4Dvs}}
\end{figure}

Figure~\ref{fig:4Dvs} shows 2D slices of these reconstructions at
values of -2.0 and 2.0 for the third and fourth backbone dihedral
angles (counted from the N-terminal end), respectively.  These
slices are compared to a reference 2D reconstruction based on data
sampled on a 2D grid of $48 \times 48$ centres in the same plane.
For large numbers of centres both methods result in acceptable
looking reconstructions. 
However, similarly to the case of reconstructing in two dimensions, only
the GPR(d) fit remains qualitatively reasonable for small numbers of
centres.  The larger gradient error of the LSRBF
reconstruction, especially for small numbers of centres as
seen in Figure~\ref{fig:4Dders}, is accompanied by prominent artefacts 
centered on the sampling points visible in Figure~\ref{fig:4Dvs}.

\onecolumn
\twocolumn

%

\section{Conclusions}\label{sec:conclusions}
In this paper we showed how Gaussian process regression can be applied to the reconstruction of relative free 
energies from molecular dynamics trajectory data---histograms, mean restraint
forces or both---obtained in umbrella sampling simulations. It is a Bayesian method that explicitly takes into 
account the prior beliefs we have about the scale and smoothness of the free energy surface and consistently deals with statistical noise in the input
data.  Our GPR-based reconstruction can use information on the probability density from histograms in GPR(h),
information on the free energy gradients in GPR(d), or both in GPR(h+d).  It thus combines aspects of many previously published methods.
GPR(h) is a density estimator, like WHAM, MBAR, vFEP, GAMUS.
GPR(d) on the other hand performs regression and integration on measurements of the free energy gradients, like UI and LSRBF.  
While MBAR, vFEP,  WHAM  and LSRBF correspond
to maximizing the likelihood, as a Bayesian method GPR maximizes the posterior and uses an explicit expression for the prior.  

We have applied GPR and other reconstruction methods to the gas phase 1 and 2-dimensional free energy surfaces of the alanine dipeptide and the 4-dimensional free energy
surface of the alanine tripeptide, using samples from molecular dynamics trajectories with quadratic restraining potentials centered
on a uniform grid of points.  We have demonstrated 
numerically that GPR leads to a small improvement in results when compared to widely used WHAM and UI methods in one dimension, and a
large improvement compared to the LSRBF and vFEP
methods in two and four dimensions.  In one dimension the variant combining histograms and gradients performs best, although in the sparse
data limit, where noise and systematic error most affect the histograms, the gradient-only variant is nearly as accurate.  
In higher dimensions, where it becomes increasingly harder to get enough data to
fill in a histogram, we use only the gradient information.  The advantage of GPR under these conditions is particularly apparent in the limits
of short sampling trajectories
(i.e.\ high statistical noise) and to some extent with sparse grids.  This robustness to noise and weak sensitivity to the squared
exponential kernel width hyperparameter are fundamental properties of the GPR approach, and are therefore expected to be transferable to free energy surface 
reconstruction in other systems.  In cases where the free energy surface is expected to be less smooth, correspondingly less smooth covariance
kernels may be used.

 We argue that GPR represents the preferable, modern and practical approach to function fitting. It  does not require any ad-hoc adjustments, rather the input parameters are intuitively meaningful and in realistic examples the results are insensitive
to a broad range of choices.
It provides meaningful error estimates with minimal extra computation.  The framework allows for the simultaneous use of 
different types of information, such as bin counts and  gradients.

The main benefit of using the GPR technique are  that shorter trajectories
can be used in mapping free energy profiles to acceptable accuracy. A reference software implementation for Gaussian process regression with
the SE kernel for function reconstruction from gradient data in an arbitrary number of dimensions 
is  freely available on www.libatoms.org.

\acknowledgement
We thank Eric Vanden-Eijnden for comments on the manuscript.  N.B. acknowledges funding for this project by
the Office of Naval Research (ONR) through the Naval Research Laboratory's basic research program.  G. C. 
acknowledges support from the Office of Naval Research under Grant No.
N000141010826 and from the European Union FP7-NMP
programme under Grant No. 229205 ``ADGLASS''.

%
%
%

\appendix
\onecolumn
\section{Derivation of the GPR(h) posterior distribution}\label{sec:convolutions}
In this Appendix we give details of the derivation of the predictive formulae stated in Eqs.~
\eqref{eq:flatprior_postmean} and \eqref{eq:flatprior_postvar} of Sec.~\ref{sec:vague}.
We start from Eq.~\eqref{eq:betalikelihood}, a Gaussian likelihood allowing for a number of unknown constants $
\mathbf{f_0}$. 
Given a set of such constants, we can simply view $\mathbf{y}-H^T\mathbf{f}_0$ as our data (which transforms Eq.~
\eqref{eq:betalikelihood} into the likelihood of the basic GPR case, Eq.~\eqref{eq:GPRlikelihood}) and thus use the 
standard predictive formulae, Eqs.~\eqref{eq:GPpostmean} and \eqref{eq:GPpostvar} to obtain $p(f(x^*)|\mathbf{y},
\mathbf{f}_0)$ as a Gaussian process with the following mean and covariance functions:
\begin{align}
\bar{f}(x^*)&=\mathbf{k}^T(x^*)K_y^{-1}\left(\mathbf{y}-H^T\mathbf{f}_0\right),\label{eq:f0postmean}\\
\cov(f(x_1^*),f(x_2^*))&=k(x_1^*,x_2^*)-\mathbf{k}^T(x_1^*)K_y^{-1}\mathbf{k}(x_2^*).\label{eq:f0postvar}
\end{align}
To get from $p(f(x^*)|\mathbf{y},\mathbf{f}_0)$ to the desired $p(f(x^*)|\mathbf{y})$, we invoke some basic relations of 
conditional and joint probabilities:
\begin{align}
p(f(x^*)|\mathbf{y})&=\int p(f(x^*),\mathbf{f}_0|\mathbf{y}) d \mathbf{f}_0\nonumber\\
&=\int p(f(x^*)|\mathbf{y},\mathbf{f}_0)p(\mathbf{f}_0|\mathbf{y}) d \mathbf{f}_0\nonumber\\
&=\frac{1}{p(\mathbf{y})}\int p(f(x^*)|\mathbf{y},\mathbf{f}_0)p(\mathbf{y}|\mathbf{f}_0)p(\mathbf{f}_0)
d \mathbf{f}_0.\label{eq:p(f|y)}
\end{align}

We are interested in the case of an uninformative (i.e.\ constant) prior distribution $p(\mathbf{f}_0)$. Consequently, 
we can take this factor outside the integral and renormalise $p(f(x^*)|\mathbf{y})$ at the end. Similarly, the marginal 
likelihood of the model, $p(\mathbf{y})=\int p(\mathbf{y}|\mathbf{f}_0)p(\mathbf{f}_0)d \mathbf{f}_0$, is simply a 
normalisation constant that we do not need to evaluate explicitly. Expressions for it can be found in Ref.~
\citenum{Rasmussen:2005ws}. The remaining factor, $p(\mathbf{y}|\mathbf{f}_0)$, is the marginal likelihood of the 
model given a specific set of values for the unknown constants $\mathbf{f}_0$,
\begin{equation}
p(\mathbf{y}|\mathbf{f}_0) = \int p(\mathbf{y} | \mathbf{f}, \mathbf{f}_0) p(\mathbf{f}| \mathbf{f}_0) d \mathbf{f}
\label{eq:marglike}
\end{equation}
This integral is a convolution of two Gaussians, the prior (Eq.~\eqref{eq:prior}) and the likelihood, Eq.~
\eqref{eq:betalikelihood}. It is a standard result that the convolution of two Gaussians is another Gaussian; the 
general case is given by
\begin{multline}
\int \exp\left[-\frac{1}{2}(\mathbf{x}_1-A\mathbf{x}_2-\boldsymbol{\mu}_1)^T\Sigma_1^{-1}(\mathbf{x}_1-A\mathbf{x}
_2-\boldsymbol{\mu}_1)\right]
\exp\left[-\frac{1}{2}(\mathbf{x}_2-\boldsymbol{\mu}_2)^T\Sigma_2^{-1}(\mathbf{x}_2-\boldsymbol{\mu}_2)\right]
d \mathbf{x}_2\\
\propto
\exp\left[-\frac{1}{2}(\mathbf{x}_1-\boldsymbol{\mu}_1-\boldsymbol{\mu}_2)^T(\Sigma_1+A\Sigma_2A^T)^{-1}
(\mathbf{x}_1-\boldsymbol{\mu}_1-A\boldsymbol{\mu}_2)\right],\label{eq:convolution}
\end{multline}
where $\mathbf{x}_1$ and $\boldsymbol{\mu}_1$ are vectors of size $n$, $\mathbf{x}_2$ and $\boldsymbol{\mu}_2$ 
vectors of size $m$ and $\Sigma_1$ and $\Sigma_2$ are symmetric positive definite matrices of sizes $n\times n$ 
and $m\times m$, respectively.

In the present case $A$ is the identity, and thus we may simply add the means (zero for the prior and $H^T\mathbf{f}
_0$ for the likelihood) and covariance matrices ($K$ and $\Sigma_y$) to obtain 
\begin{equation}
p(\mathbf{y}|\mathbf{f}_0)\propto|K_y|^{-1/2}
\exp\left[-\frac{1}{2}(\mathbf{y}-H^T\mathbf{f}_0)^TK_y^{-1}(\mathbf{y}-H^T\mathbf{f}_0)\right].\label{eq:f0marglike}
\end{equation}

Returning to Eq.~\eqref{eq:p(f|y)}, we can now see that the posterior distribution sought, $p(f(x^*)|\mathbf{y})$, is, up 
to normalisation, yet another convolution: a convolution of the posterior Gaussian process given $\mathbf{f}_0$, 
$p(f(x^*)|\mathbf{y},\mathbf{f}_0)$, and the marginal likelihood $p(\mathbf{y}|\mathbf{f}_0)$.
To apply Eq.~\eqref{eq:convolution} once more, we first need to reexpress $p(\mathbf{y|\mathbf{f}_0})$, Eq.~
\eqref{eq:f0marglike}, as a Gaussian in $\mathbf{f}_0$ rather than in $\mathbf{y}$. Expanding the product and 
completing the square yields
\begin{multline}
\log p(\mathbf{y}|\mathbf{f}_0)=
-\frac{1}{2}(\mathbf{f}_0-\bar{\mathbf{f}}_0)^T[HK_y^{-1}H^T](\mathbf{f}_0-\bar{\mathbf{f}}_0)\\
-\frac{1}{2}\mathbf{y}^TK_y^{-1}\mathbf{y}
+\frac{1}{2}\mathbf{y}^TK_y^{-1}H^T[HK_y^{-1}H^T]^{-1}HK_y^{-1}\mathbf{y}
-\frac{1}{2}\log|K_y|+\mathrm{const},\label{eq:compsq}
\end{multline}
where $\bar{\mathbf{f}}_0=[HK_y^{-1}H^T]^{-1}HK_y^{-1}\mathbf{y}$. Only the first term depends on $\mathbf{f}_0$; 
we can ignore the others as normalisation constants. We can now bring together Eqs. \eqref{eq:p(f|y)}, 
\eqref{eq:convolution}, \eqref{eq:f0postmean}, \eqref{eq:f0postvar} and \eqref{eq:compsq} to obtain Eqs.~
\eqref{eq:flatprior_postmean} and \eqref{eq:flatprior_postvar} as the mean and covariance functions for the posterior 
Gaussian process $p(f|\mathbf{y})$.

\section{Alternative formulation of GPR(h) - Learning from finite differences}\label{sec:diffs}
In this Appendix we present an alternative, but entirely equivalent, formulation of the method presented in Sec.~\ref{sec:BRUSH}. 
We can eliminate the unknown constants $\mathbf{f}_0$ by using free energy differences rather 
than absolute free energies as input data. More specifically, in each window
with $b$ bins we can use the differences of the free energies in $b-1$ bins and the remaining bin. We first spell out 
the details of this alternative approach, before demonstrating its equivalence to the view taken earlier.

The free energy differences, $\Delta\mathbf{y}$, to be used as input data can be obtained from the bin free energies, 
$\mathbf{y}$, of Sec.~\ref{sec:BRUSH} by applying a difference operator $B_{\Delta}$, i.e.
\begin{equation}
\Delta\mathbf{y}=B_{\Delta}\mathbf{y},
\end{equation}
where the (rectangular) matrix $B_{\Delta}$ has a block-diagonal structure with each $b-1 \times b$ entry block
\begin{equation}
B_{\Delta}^w=
\begin{pmatrix}
1 & 0&\cdots &0&-1\\
0& 1&\ddots&\vdots&-1\\
\vdots&\ddots&\ddots&0&\vdots\\
0&\cdots&0&1&-1\\
\end{pmatrix},
\end{equation}
i.e.\ the identity matrix extended by a column of $-1$ entries, for every window.  In this form, the final bin of each 
window is assigned to be the reference bin with respect to which the differences of the free energies of the 
remaining bins are evaluated. A different choice of reference bin would simply shift the position of the (now) 
rightmost column of $B_{\Delta}^w$. As will become clear from what follows, the choice of reference bin is entirely 
arbitrary and does not affect the result.

Using  $\Delta\mathbf{y}$ as input data in GPR we obtain the following posterior distribution:
\begin{align}
\bar{f}(x^*)&=\mathbf{k}_{\Delta}^T(x^*)K_{\Delta y}^{-1}\Delta\mathbf{y}\nonumber\\
&=\mathbf{k}^T(x^*)B_{\Delta}^T(B_{\Delta}K_yB_{\Delta}^T)^{-1}B_{\Delta}\mathbf{y},\label{eq:diffpostmean}\\
\cov(f(x_1^*),f(x_2^*))&=k(x_1^*,x_2^*)-\mathbf{k}_{\Delta}^T(x_1^*)K_{\Delta y}^{-1}\mathbf{k}_{\Delta}(x_2^*)
\nonumber\\
&=k(x_1^*,x_2^*)-\mathbf{k}^T(x_1^*)B_{\Delta}^T(B_{\Delta}K_yB_{\Delta}^T)^{-1}B_{\Delta}\mathbf{k}(x_2^*), 
\label{eq:diffpostvar}
\end{align}
where $\mathbf{k}(x^*)$, $K_y=K+\Sigma_y$ and $\mathbf{y}$ are the same as in Sec.~\ref{sec:BRUSH} and we 
have written $\mathbf{k}_{\Delta}(x^*)=B_{\Delta}\mathbf{k}(x^*)$ and $K_{\Delta y}=B_{\Delta}K_yB_{\Delta}
^T=B_{\Delta}KB_{\Delta}^T+B_{\Delta}\Sigma_yB_{\Delta}^T$.

To show the equivalence of this with respect to the posterior distribution derived in Sec.~\ref{sec:BRUSH}, we first 
note that Eqs. \eqref{eq:diffpostmean} and \eqref{eq:diffpostvar} will coincide with Eqs. \eqref{eq:flatprior_postmean} 
and \eqref{eq:flatprior_postvar} if the following holds:
\begin{equation}
B_{\Delta}^T(B_{\Delta}K_yB_{\Delta}^T)^{-1}B_{\Delta}\overset{?}{=}K_y^{-1}-K_y^{-1}H^T[HK_y^{-1}H^T]^{-1}
HK_y^{-1}.\label{eq:equivalent?}
\end{equation}
In order to prove this relation we shall find it useful to consider first the simplified case obtained by replacing 
$B_{\Delta}$ with a truncated identity matrix $I_{-w}$, obtained from the $n\times n$ identity by removing a number 
$w$ of rows from the bottom to give an $(n-w)\times n$ matrix. In the context of GPR we might interpret this as 
making use of only the first $n-w$ of $n$ data points. Multiplication of an $n\times x$ matrix from the left with $I_{-w}
$ thus removes the bottom $w$ rows of this matrix, while multiplication of an $x\times n$ matrix from the right with 
$I_{-w}^T$ removes its rightmost $w$ columns. Similarly, multiplication of an $(n-w)\times x$ matrix from the left with 
$I_{-w}^T$ will add $w$ rows of zeros and multiplication of an $x\times (n-w)$ matrix from the right with $I_{-w}$ will 
add $w$ columns of zeros. The expression $I_{-w}^T(I_{-w}K_yI_{-w}^T)^{-1}I_{-w}$ thus describes the matrix 
obtained by inverting the top-left $(n-w)\times(n-w)$ block of $K_y$ and padding the result with zeros to restore its 
size to $n\times n$. To carry on, we need to invoke a standard result (cf.\ e.g.\ Ref.~\citenum{Rasmussen:2005ws}) 
about the inverse of a partitioned matrix: Given an invertible matrix $A$ and its inverse, we may partition both 
matrices in the same way and write
\begin{equation}
A^{-1}=
\begin{pmatrix}
P&Q\\
R&S\\
\end{pmatrix}^{-1}=
\begin{pmatrix}
\tilde{P}&\tilde{Q}\\
\tilde{R}&\tilde{S}\\
\end{pmatrix},
\end{equation}
where $P$, $\tilde{P}$, $S$ and $\tilde{S}$ are square matrices. The basic result we need is
\begin{equation}
P^{-1}=\tilde{P}-\tilde{Q}\tilde{S}^{-1}\tilde{R},
\end{equation}
which we extend for our purposes to:
\begin{equation}
\begin{pmatrix}
P^{-1}&0\\
0&0\\
\end{pmatrix}=
A^{-1}-
\begin{pmatrix}
\tilde{Q}\\ \tilde{S}\\
\end{pmatrix}
\tilde{S}^{-1}
\begin{pmatrix}
\tilde{R}&\tilde{S}\\
\end{pmatrix}.\label{eq:blockinv}
\end{equation}
As explained above, the matrix $I_{-w}$ can be used to choose the top left block of a matrix. In order to use 
Eq.~\eqref{eq:blockinv} we need to define the complementary $n\times w$ matrix $H_w$, capable of selecting the 
bottom right block. This is precisely the matrix removed from the identity in order to obtain $I_{-w}$, i.e.
\begin{equation}
I=\begin{pmatrix}
I_{-w}\\ H_w
\end{pmatrix}.
\end{equation}
We can thus apply Eq.~\eqref{eq:blockinv} to our simplified problem and obtain
\begin{equation}
I_{-w}^T(I_{-w}K_yI_{-w}^T)^{-1}I_{-w}=K_y^{-1}-K_y^{-1}H_w^T\left[H_wK_y^{-1}H_w^T\right]^{-1}H_wK_y^{-1},
\label{eq:I_wK_y}
\end{equation}
which is already very close in form to Eq.~\eqref{eq:equivalent?}. To generalise this result we first note that 
$B_{\Delta}$ may be written in terms of a larger, $n\times n$ matrix $B$ as
\begin{equation}
B_{\Delta}=I_{-w}B,
\end{equation}
where the first $n-w$ rows of $B$ are the same as $B_{\Delta}$. We choose the additional rows to be the same as 
the rows of the matrix $H$, defined in Sec.~\ref{sec:BRUSH}, i.e.
\begin{equation}
H_wB=H.
\end{equation}
Crucially for what follows, these extra rows are both mutually orthogonal  as well as orthogonal to the rows of 
$B_{\Delta}$. This will allow us to write
\begin{equation}
H_w(B^{-1})^T=\diag(\mathbf{h})^{-2}H_wB=\diag(\mathbf{h})^{-2}H
\end{equation}
where the diagonal matrix $\diag(\mathbf{h})$ contains the vector norms of the rows of the matrix $H$. (For example, 
a row in H corresponding to a window with $b$ bins, will have norm $\sqrt{b}$.)

We can thus write
\begin{eqnarray}
B_{\Delta}^T(B_{\Delta}K_yB_{\Delta}^T)^{-1}B_{\Delta}&=&B^TI_{-w}^T(I_{-w}BK_yB^TI_{-w}^T)^{-1}I_{-w}B\nonumber\\
&=&B^T(B^T)^{-1}K_y^{-1}B^{-1}B\nonumber\\
&&-B^T(B^T)^{-1}K_y^{-1}B^{-1}H_w^T\left[H_w(B^T)^{-1}K_y^{-1}B^{-1}H_w^T\right]^{-1}
 H_w(B^T)^{-1}K_y^{-1}
B^{-1}B\nonumber\\
&=&K_y^{-1}-K_y^{-1}H^T\diag(\mathbf{h})^{-2}\left[\diag(\mathbf{h})^{-2}HK_y^{-1}H^T
\diag(\mathbf{h})^{-2}\right]^{-1}\diag(\mathbf{h})^{-2}HK_y^{-1}\nonumber\\
&=&K_y^{-1}-K_y^{-1}H^T[HK_y^{-1}H^T]^{-1}HK_y^{-1},
\end{eqnarray}
where the second equality is obtained by invoking Eq.~\eqref{eq:I_wK_y} with $BK_yB^T$ now replacing $K_y$.

This completes the proof of Eq.~\eqref{eq:equivalent?}. Using free energy differences in GPR is thus equivalent to 
introducing the undetermined constants $\mathbf{f}_0$. This also demonstrates that the choice of reference bin is 
indeed arbitrary and does not affect the result. Indeed, any operator $B_{\Delta}$, with rows that are mutually 
linearly independent and orthogonal to the rows of $H$, will give this same result.

\section{Derivation of the error estimates}\label{sec:errormaths}
In this Appendix we derive error estimates of the Gaussian process for the free energy differences in the periodic and nonperiodic cases.

In the first case the reconstruction repeats itself periodically and we start
by showing that the predicted mean of the reconstruction averages to zero
across the whole period, i.e. 
\begin{equation}\label{eq:0mean2pi}
\int_0^{2\pi}\bar{f}(x^*)dx^*=0.
\end{equation}
In the case of GPR(h) (Eq.~\eqref{eq:flatprior_postmean}) this expands to
\begin{equation}
\int_0^{2\pi}\bar{f}(x^*)dx^*=\left[\int_0^{2\pi}\mathbf{k}^T(x^*)dx^*\right]\left(K_y^{-1}-K_y^{-1}H^T[HK_y^{-1}
H^T]^{-1}HK_y^{-1}\right)\mathbf{y}.
\end{equation}
The vector $\mathbf{k}(x^*)$ comprises the prior covariances between the
function at $x^*$ and the location of the 
data $\mathbf{x}$, i.e.\ its elements are $(\mathbf{k}(x^*))_i=k_{2\pi}(x^*,x_i)$.
Because of the translational 
invariance of the covariance function, these integrals do not depend on the
location of the training data, i.e.
\begin{equation}
\bar{k}_{2\pi}=\int_0^{2\pi}k_{2\pi}(x^*,x_i)dx^*=\int_0^{2\pi}\tilde{k}_{2\pi}(\tau)d\tau
\end{equation}
and we have
\begin{equation}
\int_0^{2\pi}\bar{f}(x^*)dx^*=\bar{k}_{2\pi}\mathbf{1}^T\left(K_y^{-1}-K_y^{-1}H^T[HK_y^{-1}H^T]^{-1}HK_y^{-1}\right)
\mathbf{y},
\end{equation}
where $\mathbf{1}$ is a vector with all elements equal to 1. From the result
of Appendix~\ref{sec:diffs} it follows, 
however, that $\mathbf{1}^T\left(K_y^{-1}-K_y^{-1}H^T[HK_y^{-1}H^T]^{-1}HK_y^{-1}\right)$
is equal to the zero 
vector and so we obtain Eq.~\eqref{eq:0mean2pi}.
In the case of GPR(d) (Eq.~\eqref{eq:derpostmean}), we get
\begin{equation}
\int_0^{2\pi}\bar{f}(x^*)dx^*=\left[\int_0^{2\pi}\mathbf{k}_{f,f'}^T(x^*)dx^*\right](K_{f',f'}(\mathbf{x}')+
\Sigma_{y'})^{-1}\mathbf{y'}.
\end{equation}
The vector $\mathbf{k}_{f,f'}(x^*)$ has elements (Eq.~\eqref{eq:kff'})
\begin{equation}\label{eq:diffk2pi}
k_{f,f'}(x^*,x'_i)=\frac{\partial}{\partial x'_i} k_{2\pi}(x^*,x'_i)=-\frac{\partial}{\partial
x^*} k_{2\pi}(x^*,x'_i),
\end{equation}
where the second equality is a direct consequence of the translational invariance
of the covariance function. 
Integrating these covariance functions over $x^*$ thus gives 0 for all $x'_i$
and Eq.~\eqref{eq:0mean2pi} 
immediately follows. The above arguments are virtually the same in the case
of a covariance function of our second 
type and we get
\begin{equation}
\lim_{\Delta\rightarrow \infty} \quad \frac{1}{2\Delta}\int_{-\Delta}^{\Delta}\bar{f}(x^*)dx^*=0
\end{equation}
in this case.

To obtain error bars on the reconstructed free energy profile, we need to
find expressions for
\begin{equation}
\var\left[f(x^*)-\frac{1}{2\pi}\int_{0}^{2\pi}f(x^*_1)dx^*_1\right]
\end{equation}
and
\begin{equation}
\var\left[f(x^*)-\lim_{\Delta\rightarrow\infty}\frac{1}{2\Delta}\int_{-\Delta}^{\Delta}f(x_1^*)dx_1^*\right],
\end{equation}
respectively, depending on the covariance function used.  Note the presence of
the full posterior $f$ in the above integrals, 
rather than the posterior mean $\bar{f}$. 
We are now going to show that suitable error estimates are provided by 
\begin{equation}
\var\left[f(x^*)-\frac{1}{2\pi}\int_{0}^{2\pi}f(x_1^*)dx_1^*\right]=\var[f(x^*)]-\frac{\bar{k}_{2\pi}}{2\pi}
\label{eq:errorbars}
\end{equation}
for the periodic case, and
\begin{equation}
\var\left[f(x^*)-\lim_{\Delta\rightarrow\infty}\frac{1}{2\Delta}\int_{-\Delta}^{\Delta}f(x_1^*)dx_1^*\right]=\var[f(x^*)]
\label{eq:errorbars_infty}
\end{equation}
for the non-periodic case.

In the periodic case we obtain,
\begin{multline}
\var\left[f(x^*)-\frac{1}{2\pi}\int_{0}^{2\pi}f(x_1^*)dx_1^*\right]=\\
\var[f(x^*)]+\var\left[\frac{1}{2\pi}\int_{0}^{2\pi}f(x_1^*)dx_1^*\right]
-2\cov\left[f(x^*),\frac{1}{2\pi}\int_{0}^{2\pi}f(x_1^*)dx_1^*\right]\\
=\var[f(x^*)]+\frac{1}{(2\pi)^2}\int_{0}^{2\pi}\int_{0}^{2\pi}\cov[f(x_1^*),f(x_2^*)]dx_1^*dx_2^*
-\frac{2}{2\pi}\int_{0}^{2\pi}\cov[f(x^*),f(x_1^*)]dx_1^*.\label{eq:varexpansion}\\
\end{multline}
In the case of GPR(h) we can expand the double integral as follows (cf.\ Eq.~\eqref{eq:flatprior_postvar}):
\begin{multline}
\int_{0}^{2\pi}\int_{0}^{2\pi}\cov[f(x_1^*),f(x_2^*)]dx_1^*dx_2^*=\\
\int_{0}^{2\pi}\int_{0}^{2\pi}k_{2\pi}(x_1^*,x_2^*)dx_1^*dx_2^*
-\sum_{ij}\kappa_{ij}\left(\int_{0}^{2\pi}k_{2\pi}(x_i,x_1^*)dx_1^*\right)\left(\int_{0}^{2\pi}k_{2\pi}(x_j,x_2^*)dx_2^*
\right)=\\
2\pi\bar{k}_{2\pi}-\bar{k}_{2\pi}^2\mathbf{1}^T\kappa\mathbf{1}=2\pi\bar{k}_{2\pi},
\end{multline}
where $\kappa$ denotes the matrix $K_y^{-1}-K_y^{-1}H^T[HK_y^{-1}H^T]^{-1}HK_y^{-1}$ and we have again used 
the result that $\mathbf{1}^T\kappa=\mathbf{0}$.
Similarly,
\begin{equation}
\int_{0}^{2\pi}\cov[f(x^*),f(x_1^*)]dx_1^*=\bar{k}_{2\pi},
\end{equation}
and we obtain Eq.~\eqref{eq:errorbars},
\begin{equation*}
\var\left[f(x^*)-\frac{1}{2\pi}\int_{0}^{2\pi}f(x_1^*)dx_1^*\right]=\var[f(x^*)]-\frac{\bar{k}_{2\pi}}{2\pi}.
\end{equation*}
It is straightforward to show that this result also holds for GPR(d). One simply expands the integrals of Eq.~
\eqref{eq:varexpansion} using Eq.~\eqref{eq:derpostvar} and all integrals involving differentiated covariance 
functions evaluate to zero (cf.\ Eq.~\eqref{eq:diffk2pi}), leaving Eq.~\eqref{eq:errorbars}.

Again, a very similar line of reasoning can be employed for our second type of covariance function, though the 
diverging denominator provides an obvious shortcut in this case. It is also because of this diverging denominator 
(compared to the finite value of $2\pi$ in the first case) that the final result, Eq.~\eqref{eq:errorbars_infty}, is even 
simpler:
\begin{equation*}
\var\left[f(x^*)-\lim_{\Delta\rightarrow\infty}\frac{1}{2\Delta}\int_{-\Delta}^{\Delta}f(x_1^*)dx_1^*\right]=\var[f(x^*)].
\end{equation*}

\twocolumn

\bibliography{BRUSH}

\newpage

\Large for Table of Contents use only

\vspace*{0.5in}

{\includegraphics[width=8.9cm]{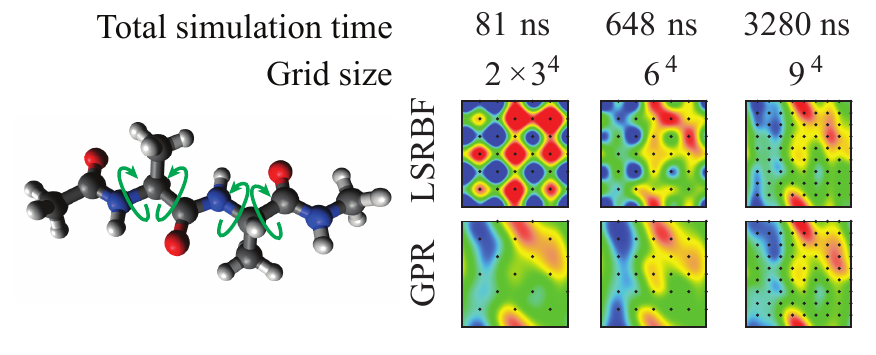}}
\end{document}